\documentclass[12pt,a4paper]{article}
\usepackage[T2A]{fontenc}
\usepackage[cp1251]{inputenc}
\usepackage{cite}
\usepackage{mathtext}
\usepackage{amssymb,amsthm,amsmath}
\usepackage{array}
\usepackage[dvips]{epsfig}

\begin{document}
\author{\bf Yuri A. Markov$\!\,$\thanks{e-mail:markov@icc.ru}\,,\,
Margaret A. Markova$^*$,\\
\bf and Alexey A. Shishmarev$\!\,$\thanks{e-mail:a.a.shishmarev@mail.ru}}
\title{The equations of motion for a classical color particle\\
in background non-Abelian bosonic and\\ fermionic fields.}
\date{\it Institute for System Dynamics\\
and Control Theory, Siberian Branch\\
of Academy of Sciences of Russia,\\
P.O. Box 1233, 664033 Irkutsk, Russia}

\thispagestyle{empty}
\maketitle{}

%%%% Making dual numbered equations %%%%%%%%%%%%

\def\theequation{\arabic{section}.\arabic{equation}}

%\newcounter{equation}[section]
%\renewcommand{\theequation}{\thesection.\arabic{equation}}
\[
{\bf Abstract}
\]
{Based on the most general principles of reality, gauge and reparametrization invariance, a problem of constructing the action
describing dynamics of a classical color-charged particle interac\-ting with background non-Abelian gauge and fermion fields
is considered. The cases of the linear and quadratic dependence of a Lagrangian on a background Grassmann fermion field are discussed.
It is shown that in both cases in general there exists an infinite number of interaction terms, which should be included in
the Lagrangian in question. Employing a simple iteration scheme, examples of the construction of the first few gauge-covariant
currents and sources induced by a moving particle with non-Abelian charge are given. It is found that these quantities,
by a suitable choice of parameters, exactly reproduce additional currents and sources  previously obtained in
[\,Yu.A. Markov, M.A. Markova, Nucl.\,Phys.\,A {\bf 784} (2007) 443] on the basis of heuristic considerations.}

%{\sl PACS:} 12.38.Mh, 24.85.+p, 11.15.Kc

\newpage

\section{Introduction}
\setcounter{equation}{0}

The formulation of Lagrangian and Hamiltonian descriptions of the dynamics of (pseudo)clas\-sical color particles interacting with a background Yang-Mills field
has been suggested more then 30 years ago in two fundamental papers by Barducci {\it et al} \cite{barducci_1977} and Balachandran {\it et al} \cite{balachandran_1977}.
In the present work, we have attempted to extend the approaches developed in the papers to the case of the presence of a background `non-Abelian' fermion field in the system along with a background non-Abelian gauge field. As a practical guidance, at least at the first stage of the solution to the stated problem, we use the general principles, which were first  formulated more precisely in \cite{balachandran_1977}, namely, the desired action should satisfy the following conditions:
\begin{itemize}
\item[(1)]
it must be real up to a total time derivative;
\item[(2)]
it must be invariant under the coordinate transformation of the parameter of $\mbox{integration}\;\tau$, i.e. under the replacement
\begin{equation}
\tau \rightarrow{\tau}^{\prime} = f(\tau);
\label{eq:1q}
\end{equation}
\item[(3)]
it must be invariant under the gauge transformations.
\end{itemize}
Besides, in \cite{balachandran_1977} the requirement of the consistency with the Wong equation \cite{wong_1970} was added. In the papers \cite{balachandran_1977, barducci_1977}
a simple action for a particle with non-Abelian charge moving in a gauge field background satisfying all the above-listed requirements has been independently suggested, namely,
\[
S=\!\int\!\!L\,d\tau\,,
\]
where the Lagrangian reads
\begin{equation}
L=-m\sqrt{\dot{x}^{\mu}\dot{x}_{\mu}}+i\hspace{0.02cm}\theta^{\dagger{i\!}}D^{ij}\theta^{j}.
\label{eq:1w}
\end{equation}
Here, $D^{ij}=\delta^{ij}\partial/\partial\tau+ig\hspace{0.02cm}\dot{x}^{\mu\!}A^{a}_{\mu}(t^{a})^{ij}$ is the covariant derivative along the direction of motion; $\theta^{\dagger{i}}$ and $\theta^i$ are a set of Grassmann variables belonging to the fundamental representation of the $SU(N_c)$ color group, i.e. $i=1,\ldots,N_c$. The equations of motion are
\begin{equation}
m\,\frac{d}{d\tau}\!\left(\frac{\dot{x}^{\mu}}{(\dot{x}^{\nu}\dot{x}_{\nu})^{1/2}}\right)=
g\hspace{0.03cm}Q^aF^{a}_{\mu\nu\,}\dot{x}^{\nu},
\label{eq:1e}
\end{equation}
\begin{equation}
\begin{split}
&\frac{d\theta^i(\tau)}{d\tau}\,+\,ig\hspace{0.025cm}\dot{x}^{\mu}(\tau)A_{\mu}^{a}(x)(t^a)^{ij}\theta^j(\tau)=0,\\
&\frac{d\theta^{\dagger{i}}(\tau)}{d\tau}\,-ig\hspace{0.025cm}\dot{x}^{\mu}(\tau)A_{\mu}^{a}(x)\theta^{\dagger{j}}(\tau)(t^a)^{ji}=0,
\end{split}
\label{eq:1r}
\end{equation}
where in the first equation we have set $Q^a\equiv\theta^{\dagger{i}}(t^a)^{ij}\theta^{j}$. Making use of the equations of motion for the Grassmann color charges $\theta^i$ and $\theta^{\dagger{i}}$, it is easily to see that the classical commuting color charge $Q=(Q^a),\,a=1,\ldots ,N_c^2-1$, satisfies Wong's equation
\begin{equation}
\frac{dQ^a(\tau)}{d\tau}\,+\,ig\hspace{0.025cm}\dot{x}^{\mu}(\tau)A^b_{\mu}(x)(T^b)^{ac}Q^c(\tau)=0.
\label{eq:1t}
\end{equation}
\indent
The equations of motion for Grassmann and usual color charges, (\ref{eq:1r}) and (\ref{eq:1t}), have been used in our two papers \cite{markov_NPA_2007, markov_IJMPA_2010} in the study of the scattering processes of hard color-charged particles off soft gluon and quark-antiquark fields. These soft fields are induced by thermal fluctuations in a hot quark-gluon plasma. Putting into consideration the Grassmann color charges of a hard particle on an equal footing with the usual color charge\footnote{\,We may regard these new Grassmann variables as the anticommuting degrees of freedom of the `total' color (super?)charge for the particle.}, enables us to introduce so-called color (Grassmann valued) source of a spin-1/2 hard particle along with classical color current. We add this Grassmann source to the right-hand side of the Dirac equation for soft fermion field just as we add the usual color current of the hard particle to the right-hand side of the Yang-Mills equation for soft boson field \cite{markov_AOP_2004_2005}. This allowed us to obtain a closed self-consistent description of nonlinear interaction dynamics of soft and hard excitations of the hot QCD plasma both Fermi and Bose statistics (within the framework of semiclassical approximation).\\
\indent
However, as was shown in \cite{markov_NPA_2007}, the equations for Grassmann charges (\ref{eq:1r}) and the Wong equation (\ref{eq:1t}) as they stand in the original works
\cite{balachandran_1977, barducci_1977,wong_1970} are insufficient to obtain complete and gauge invariant expressions for the matrix elements of some scattering processes. The reason for this
lies in the fact that the equations were obtained under assumption that there exists only (regular and/or stochastic) background gluon field $A_{\mu}^a(x)$ in the system. It is pertinent
at this point to note that in the presence of only background gauge field, putting into consideration Grassmann color charges $\theta^{\dagger{i}}$ and $\theta^i$ gives merely the possibility of an elegant Lagrangian (or Hamiltonian) formulation. However, in actual dynamics of a color particle, the presence of the Grassmann charges is not revealed in any way. This is linked with the fact that only the bilinear (i.e. Grassmann-even) combination $\theta^{\dagger}t^a\theta\,(\equiv\!Q^a)$ representing itself commuting classical color charge appears in the equation of motion for the position $x_{\mu}$, Eq.\,(\ref{eq:1e}), and also in the expression for the color current $j_{\mu}^a(x)$ induced by this particle, Eq.\,(\ref{eq:2p}). The situation can qualitatively change only if the system will be subjected to background non-Abelian fermionic field, which as though `splits' the combination $\theta^{\dagger}t^a\theta$ into two independent (Grassmann-odd) parts. Here, the necessity of introducing Grassmann color charges as dynamical variables enjoying full rights should be manifested in full.\\
\indent
In the work \cite{markov_NPA_2007}, we proposed a minimal extension of equations (\ref{eq:1r}) to the case of the presence of soft (stochastic) fermion fields $\bar{\Psi}^i_{\alpha}(x)$ and $\Psi^i_{\alpha}(x)$ in the system. Thus instead of (\ref{eq:1r}) we have the following generalized equations for the Grassman color charges:
\begin{equation}
\begin{split}
&\frac{d\theta^i(t)}{dt} + ig\hspace{0.025cm}v^{\mu\!}A^a_{\mu}(t,{\bf v}t)(t^a)^{ij}
\theta^j(t)
+ig\bigl(\bar{\chi}_{\alpha}\Psi_{\alpha}^i(t,{\bf v}t)\bigr)=0,
\quad\;\;
\left.\theta^i_0=\theta^i(t)\right|_{\,t=t_0},\\
&\frac{d\theta^{\dagger i}(t)}{dt} - ig\hspace{0.025cm}v^{\mu\!}A^a_{\mu}(t,{\bf v}t)
\theta^{\dagger j}(t)(t^a)^{ji}
-ig\bigl(\bar{\Psi}_{\alpha}^i(t,{\bf v}t){\chi}_{\alpha}\bigr)=0,
\quad
\left.\theta^{\dagger i}_0=\theta^{\dagger i}(t)\right|_{\,t=t_0},
\end{split}
\label{eq:1y}
\end{equation}
and similar, instead of the Wong equation (\ref{eq:1t}) now we have the generalized Wong equation
\[
\frac{dQ^a(t)}{dt} + ig\hspace{0.025cm}v^{\mu\!}A^b_{\mu}(t,{\bf v}t)(T^b)^{ac}Q^c(t)
\hspace{7.65cm}
\]
\begin{equation}
\hspace{3cm}
+\;ig\Bigl[\,\theta^{\dagger j}(t)(t^a)^{ji}
\bigl(\bar{\chi}_{\alpha}\Psi_{\alpha}^i(t,{\bf v}t)\bigr)-
\bigl(\bar{\Psi}_{\alpha}^i(t,{\bf v}t){\chi}_{\alpha}\bigr)
(t^a)^{ij}\theta^j(t)\Bigr]\!=\!0
\label{eq:1u}
\end{equation}
with the initial condition $\left.Q^a_0 = Q^a(t)\right|_{\,t=t_0}$. Here, $v^{\mu}\equiv(1,{\bf v})$, $(T^a)^{bc}\equiv-if^{abc}$, and $t$ is the coordinate time. Furthermore, $\chi_{\alpha}$ is a $c$\,-number spinor describing spin state of a particle. In the papers \cite{markov_NPA_2007, markov_IJMPA_2010}, this spinor was considered as independent of time. The minimal information about this spinor we shall need in the present work, is its connection with the density matrix for completely unpolarized state of the particle. Let us define a polarization matrix $\varrho=(\varrho_{\alpha\beta})$ for the spin-1/2 particle  such that in a pure state it is reduced to a product
\[
\varrho_{\alpha\beta}=\chi_{\alpha}\bar{\chi}_{\beta}.
\]
In the paper \cite{markov_NPA_2007}, it has been shown that for the case of a completely unpolarized state, this matrix should have the following form:
\begin{equation}
\varrho=\varrho(E,{\bf v})=\frac{1}{2E}\,\varrho({\bf v}),
\label{eq:1i}
\end{equation}
where
\[
\varrho({\bf v})=\frac{1}{2}\;(v\cdot\gamma).
\]
The multiplier $1/2E$ is chosen for reasons of dimension\footnote{\,Since in \cite{markov_NPA_2007} we were interested in the case of ultrarelativistic particles, in the expression for
$\varrho(E,{\bf v})$ the term proportional to $m/E$ was dropped.}. It is to be noted specially that as distinct from (\ref{eq:1r}) and (\ref{eq:1t}), the equations (\ref{eq:1y}) and (\ref{eq:1u}) are written in the coordinate-time representation, and the background fields $A_{\mu}^a(x)$, $\bar{\Psi}^i_{\alpha}(x)$ and $\Psi^i_{\alpha}(x)$ entering into them are defined on the straight path ${\bf x}={\bf v}t$ (i.e. here, we neglect by a change of particle trajectory). It is not difficult to write down the Lagrangian whose variation would give equations (\ref{eq:1y})
\begin{equation}
L=i\hspace{0.025cm}\theta^{\dagger\, i}\dot{\theta^i}-
g\hspace{0.025cm}v^{\mu\!}A^a_{\mu}\hspace{0.03cm}\theta^{\dagger{i}}(t^a)^{ij}\theta^j-
g\Bigl\{\theta^{\dagger{i}}
(\bar{\chi}_{\alpha}\Psi_{\alpha}^i)+
(\bar{\Psi}_{\alpha}^i{\chi}_{\alpha})\theta^i\Bigr\}.
\label{eq:1o}
\end{equation}
\indent
Aside from the generalized equations of motion for the color charges in \cite{markov_NPA_2007, markov_IJMPA_2010}, new gauge-covariant additional color currents and sources
generated by a moving color particle, which should be added to the right-hand side of the proper field equations, have been suggested. In this case only we are able to calculate complete and gauge-covariant expressions for effective currents and sources generating the scattering processes of soft quark excitations off hard thermal particles in a hot QCD plasma. For convenience of the further references, the list of all additional color currents and sources obtained in the papers \cite{markov_NPA_2007, markov_IJMPA_2010} is given in $\mbox{Appendix\;A}.$\\
\indent
Unfortunately, these additional color currents and sources have been derived in the works \cite{markov_NPA_2007, markov_IJMPA_2010} mainly from heuristic reasoning, practically without any connection with dynamical equations (\ref{eq:1y}) and (\ref{eq:1u}). Here, we would like to have some systematic procedu\-re, which would enable us to obtain these quantities to any order in the coupling constant. One of the purposes of the present work is to suggest an algorithm of calculation of all gauge-covariant additional color currents and sources with any degree of accuracy in powers of the background fermion fields $\bar{\Psi}^i_{\alpha}(x)$, $\Psi^i_{\alpha}(x)$ and initial values $Q^a_0$, $\theta^{\dagger{i}}_0$ and $\theta^i_0$ of the color charges, based {\it exclusively} on the equations of motion for the Grassmann color charges. The extended Lagrangian (\ref{eq:1o}) does not give us such a possibility. This circumstance can be considered as an indication that some terms involving the background fermionic field are overlooked in this Lagrangian and restoring these terms is our first task.\\
\indent
It is hoped that research of the problem of motion for a point particle (which can be regarded as string length-zero limits) in the background fermionic field will make possible to better understanding, at least at a qualitative level, a similar motion of much more complicated object such as string. Special interest in research of motion of a spinning string in background fermionic fields exists already for many years beginning with pioneer works by Callan, Friedan {\it et al} \cite{callan_1985, friedan_1986}, and ending with more recent studies devoted to strings in Ramond-Ramond backgrounds (see, e.g., \cite{berenstein_1999_2000, hassan_2000}).\\
\indent
At the end one general point need to be made. Throughout this work we use classical Grassmann-valued charges and external fermion fields. It is necessary to give a little motivation why we use anticommuting variables rather than conventional commuting (complex) ones. Two approaches to the description of internal color symmetry, by using commuting and anticommuting color charges, have been discussed by Balachandran {\it et al} in \cite{balachandran_1977}. The principle difference between these two approaches arises at quantization of the classical models. Starting from commuting non-Abelian variables, the quantized non-Abelian charge can take arbitrarily large quantum numbers, while beginning with anticommuting variables, only a finite number of quantum numbers for the non-Abelian charge of a particle are obtained. From the physical point of view, it is clear that finite-dimensional representations of the internal color symmetry group are certainly preferable to infinite-dimensional ones (for this reason in \cite{barducci_1977} the case of commuting isospin variables has not been discussed at all). By virtue of the above-mentioned reason, if one keeps in mind further applications and also the problem of quantization of the model considered in the present work, we have preferred from the outset to work with anticommuting dynamical variables $\theta^{\dagger i}$ and $\theta^i$. In addition we can say that it is precisely these Grassmann variables that arise within the framework of the worldline path integral representation for the effective QCD action when the internal color degrees of freedom are expressed in terms of wordline fermions. This question is discussed in Conclusion in more detail.\\
\indent
Besides, the need of using Grassmann-valued color charges and background fields is connected with the fact that in computing the probabilities for various scattering processes involving hard and soft fermionic excitations in a hot quark-gluon plasma, we get automatically gauge-invariance expressions. In the opposite case, we get the incorrect signs of different terms in these probabilities resulting in violation of gauge symmetry. This was shown by straightforward calculations for the scattering processes with soft fermion excitations only \cite{markov_PRD_2001, markov_NPA_2006} as well as for the scattering processes with hard particles \cite{markov_NPA_2007, markov_IJMPA_2010}.\\
\indent
Last but not least is concerned with the possibility of keeping track of a connection of the representations given here with the problem of motion of a spinning string in fermionic background fields. In the string theory, these background fields by initial construction are anticommuting ones, and thereby we have little choice.\\
\indent
%%%%%%%%%%%%%%%%%%%%%%%%%%%%%%%%%%%%%%%%%%%%%%%%%%%%%%%%%%%%%%%%%%%%%%%%%%%%%%%%%%%%%%%%%%%%%%%%%%%%%%%%%%%%%%%%%%%%%%%%%%%%%%%%%%%%%%%%%%%%%%%%%%%%%%%%%%%%%%%%%%%%%%%%%%%%%%%%%%%%%%%%%%%%%%%%%%%%%%%
The paper is organized as follows. In Section 2, as the first example, the most simple and at the same time sufficiently meaningful extension of the Lagrangian (\ref{eq:1w}) to the case
of the presence of a background non-Abelian fermion field in the system along with a non-Abelian gauge field is suggested. All of the equations of motion, and also the color current and source generated by a moving color-charged particle, are written down in an explicit form. Section 3 is devoted to a discussion of a possibility of taking into consideration a set of the real Grassmann charges belonging to the adjoint representation of the $SU(N_c)$ group. In Section 4, based upon the requirements of gauge invariance and reality, a question of the most general structure of the Lagrangian linear depending on the external fermionic fields $\bar{\Psi}^i_{\alpha}(x)$ and $\Psi^i_{\alpha}(x)$ is taken up. It is shown that within the framework of these general requirements, one can determine an infinite number of contributions to the interaction Lagrangian, containing the strength tensor $F_{\mu\nu}^a(x)$ to an arbitrary power. The following Section 5 is concerned with a similar analysis for the case of the quadratic dependence of the Lagrangian on $\bar{\Psi}^i_{\alpha}(x)$ and $\Psi^i_{\alpha}(x)$. It has been found that in this case the interaction Lagrangian possesses richer and more varied structure, than it was in the case of the linear dependence. It is demonstrated that in principle it also can include an infinite number of interaction terms. Furthermore, in Section 6, a simple iteration scheme for the construction of the consequence of more and more becoming complicated gauge-covariant currents and sources induced by a moving color particle is suggested. It is shown that fixing one arbitrary parameter only enables us to reproduce exactly additional currents and sources obtained previously, which are listed in Appendix A. In Appendix B, an explicit form of the one-loop effective QCD action deduced within the second order formalism for fermions is given.\\
\indent
In the concluding section we briefly discuss a question of rigorous proof of the obtained results within the framework of the worldline path integral approach.
%%%%%%%%%%%%%%%%%%%%%%%%%%%%%%%%%%%%%%%%%%%%%%%%%%%%%%%%%%%%%%%%%%%%%%%%%%%%%%%%%%%%%%%%%%%%%%%%%%%%%%%%%%%%%%%%%%%%%%%%%%%%%%%%%%%%%%%%%%%%%%%%%%%%%%%%%%%%%%%%%%%%%%%%%%%%%%%%%%%%%%%%%%%%%%%%%%%%%%%%%

\section{The simplest model Lagrangian for color charge dyna\-mics in a background fermion field}
\setcounter{equation}{0}

We consider ${\rm SU}(N_c)$ gauge theory, use the metric $g^{\mu \nu} = {\rm diag}(1,-1,-1,-1)$, choose units such that $c=1$ and note $x=(x_0,{\bf x})$ etc. The color indices for the adjoint representation $a,b, \ldots$ run from 1 to $N_c^2-1$, while those for the fundamental representation $i,j, \ldots$ run from 1 to $N_c$. The Greek indices $\alpha, \beta, \ldots$ for the spinor representation run from 1 to 4.\\
\indent
Let us assume that a point-like classical color-charged particle propagates in background non-Abelian bosonic and fermionic fields. We state that the dynamics of this particle within the simplest
of possible models can be correctly described by the following action:
\begin{equation}
S=\!\int\!\!L\,d\tau,
\label{eq:2q}
\end{equation}
where the Lagrangian $L$ for such a particle is defined by the expression
\begin{equation}
L\equiv L_{\theta}=-\frac{1}{2e}\,\dot{x}^{\mu}\dot{x}_{\mu} - \frac{e}{2}\,m^2 + i\theta^{\dagger{i}}D^{ij}\theta^{j}
\label{eq:2w}
\end{equation}
\[
-\,\frac{e}{\sqrt{2}}\;g\Bigl\{\theta^{\dagger{i}}\bigl(\bar\psi_{\alpha}\Psi^{i}_{\alpha}\bigr)+\bigl(\bar\Psi^{i}_{\alpha}\psi_{\alpha}\bigr)\theta^{i}\Bigr\}+
\frac{e}{\sqrt{2}}\,gQ^a\Bigl\{f_{0\,}\theta^{\dagger{i}}(t^{a})^{ij}\bigl(\bar\psi_{\alpha}\Psi^{j}_{\alpha}\bigr)
+f_0^{\ast}\bigl(\bar\Psi^{j}_{\alpha}\psi_{\alpha}\bigr)(t^{a})^{ji}\theta^{i}\Bigr\}.
\]
Here, $\psi_{\alpha}$ is a $c$\,-number spinor describing the spin degree of freedom of the particle, $e$ is the (one-dimensional) vierbein field and $f_0$ is some (complex) scalar gauge-invariant function. The point of entering the function $f_0$ into the Lagrangian (\ref{eq:2w}) will be discussed in detail in $\mbox{Section\,4}.$ Once again let us recall that the color charge $Q^a$ is defined by the following expression:
\begin{equation}
Q^a\!\equiv\theta^{\dagger{i}}(t^a)^{ij}\theta^{j}.
\label{eq:2e}
\end{equation}
An important difference of the Lagrangian (\ref{eq:2w}) from the Lagrangian (\ref{eq:1o}) is the presence of the last term proportional to the commuting color charge (\ref{eq:2e}).\\
\indent
The action (\ref{eq:2q}) is reparametrization invariant if a change of the parameter of $\mbox{integration}\,\tau$, Eq.\,(\ref{eq:1q}), is accompanied by variable transformations
\[
\begin{split}
&e \rightarrow e^{\prime}=e\,\frac{d\tau}{d\tau^{\prime}},\\
&x^{\mu},\;\psi_{\alpha},\;f_0\quad{\rm are\;unchanged}.
\end{split}
\]
Furthermore, it is quite clear that the Lagrangian is real, and it can be shown that it is invariant with respect to the infinitesimal gauge transformations:
\begin{equation}
\begin{split}
\Psi^{i}_{\alpha}&\rightarrow\Psi^{i}_{\alpha}+ig\Lambda^{a}(t^{a})^{ij}\Psi^{j}_{\alpha},\\
\bar\Psi^{i}_{\alpha}&\rightarrow\bar\Psi^{i}_{\alpha}-ig\Lambda^{a}\bar{\Psi}^{j}_{\alpha}(t^{a})^{ji},\\
\theta^i\,&\rightarrow\;\theta^i+ig\Lambda^a(t^a)^{ij}\theta^j,\\
\theta^{\dagger{}i}&\rightarrow\theta^{\dagger{}i}-ig\Lambda^a\theta^{\dagger{}j}(t^a)^{ji},\\
Q^a&\rightarrow{Q}^a-gf^{abc}\Lambda^{b}Q^c,\\
A^{a}_{\mu}&\rightarrow{A}^{a}_{\mu}-gf^{abc}\Lambda^{b}{A}^{c}_{\mu}-\partial_{\mu}\Lambda^{a},
\label{eq:2r}
\end{split}
\end{equation}
where $\Lambda^a$ is a parameter of the transformations. For simplicity throughout our work, we neglect a change of spin state of the particle, i.e.\,we believe $\psi_{\alpha}$ to be a spinor independent of the parameter $\tau$. The account for the spin degree of freedom in the general dynamics of the particle will be considered in our next paper \cite{part_II}. In particular, there it will be shown how one can connect the $c$\,-number spinors $\bar{\psi}_{\alpha}$ and $\psi_{\alpha}$ with the pseudovector and pseudoscalar dynamical variables $\xi^{\mu}, \, \mu = 0, 1, 2, 3$ and $\xi^5$ commonly used in a description of the spin degree of freedom of massive spinning particles, and which in turn are elements of the Grassmann algebra \cite{berezin_1975, berezin_1977}.\\
\indent
Varying the variable $e$ gives the constraint equation
\[
\frac{\dot{x}^2}{e^2}\,-\,m^2-\,
\sqrt{2}\,g\Bigl\{\theta^{\dagger{i}}\bigl(\bar\psi_{\alpha}\Psi^{i}_{\alpha}\bigr)+\bigl(\bar\Psi^{i}_{\alpha}\psi_{\alpha}\bigr)\theta^{i}\Bigr\}
+\,\sqrt{2}\,g\hspace{0.025cm}Q^a\Bigl\{f_{0\,}\theta^{\dagger{i}}(t^{a})^{ij}\bigl(\bar\psi_{\alpha}\Psi^{j}_{\alpha}\bigr)
+f_0^{\ast}\bigl(\bar\Psi^{j}_{\alpha}\psi_{\alpha}\bigr)(t^{a})^{ji}\theta^{i}\Bigr\}\!=0.
\]
For the remainder of our work, we choose a parametrization in which $e=1/m$. Furthermore, varying the Lagrangian (\ref{eq:2w}) with respect to Grassmann color charge $\theta^{\dagger{i}}$, we obtain the evolution equation for $\theta^i$ in the following form:
\begin{equation}
\frac{d\theta^i(\tau)}{d\tau}
+ ig\hspace{0.01cm}\dot{x}^{\mu}(\tau)A_{\mu}^{a}(x)(t^a)^{ij}\theta^j(\tau)
\label{eq:2t}
\end{equation}
\[
+\,\frac{ig}{\sqrt{2}\,m}\,\bigl(\bar{\psi}_{\alpha}\Psi^{i}_{\alpha}(x)\bigr)
-\frac{ig}{\sqrt{2}\,m}\,f_{0\,}Q^a(\tau)(t^a)^{ij}\bigl(\bar{\psi}_{\alpha}\Psi^{j}_{\alpha}(x)\bigr)
\]
\[
-\,\frac{ig}{\sqrt{2}\,m}\,(t^a)^{ij}\theta^{j}(\tau)\,\Bigl\{\!f_{0\,}\theta^{\dagger{l}}(\tau)(t^{a})^{lk}\bigl(\bar\psi_{\alpha}\Psi^{k}_{\alpha}(x)\bigr)
+f_{0}^{\ast}\bigl(\bar\Psi^{k}_{\alpha}(x)\psi_{\alpha}\bigr)(t^{a})^{kl}\theta^{l}(\tau)\Bigr\}=0.
\]
Correspondingly one can define the equation for conjugate charge  $\theta^{\dagger{i}}$:
\begin{equation}
\frac{d\theta^{{\dagger}i}(\tau)}{d\tau}
-ig\hspace{0.01cm}\dot{x}^{\mu}(\tau)A_{\mu}^{a}(x)\theta^{{\dagger}j}(\tau)(t^a)^{ji}
\label{eq:2y}
\end{equation}
\[
-\,\frac{ig}{\sqrt{2}\,m}\,\bigl(\bar\Psi^{i}_{\alpha}(x){\psi}_{\alpha}\bigr)
+\frac{ig}{\sqrt{2}\,m}\,f_0^{\ast}Q^a(\tau)\bigl(\bar\Psi^{j}_{\alpha}(x){\psi}_{\alpha}\bigr)(t^a)^{ji}
\]
\[
+\,\frac{ig}{\sqrt{2}\,m}\,\theta^{\dagger{j}}(\tau)(t^a)^{ji}\Bigl\{\!f_{0\,}\theta^{\dagger{l}}(\tau)(t^{a})^{lk}\bigl(\bar\psi_{\alpha}\Psi^{k}_{\alpha}(x)\bigr)
+f_0^{\ast}\bigl(\bar\Psi^{k}_{\alpha}(x)\psi_{\alpha}\bigr)(t^{a})^{kl}\theta^{l}(\tau)\Bigr\}=0,
\]
where the charge $Q^a$ in view of (\ref{eq:2e}) obeys the equation:
\begin{equation}
\frac{dQ^a(\tau)}{d\tau} + ig\hspace{0.01cm}\dot{x}^{\mu}(\tau)A^b_{\mu}(x)(T^b)^{ac}Q^c(\tau)
\label{eq:2u}
\end{equation}
\[
\begin{split}
+\,\frac{ig}{\sqrt{2}\,m}\,&\Bigl\{\theta^{\dagger i}(\tau)(t^a)^{ij}
\bigl(\bar{\psi}_{\alpha}\Psi_{\alpha}^j(x)\bigr)-
\bigl(\bar{\Psi}_{\alpha}^j(x){\psi}_{\alpha}\bigr)
(t^a)^{ji}\theta^i(\tau)\Bigr\}\\
-\,\frac{ig}{\sqrt{2}\,m}\,Q^b(\tau)&\Bigl\{\!f_{0\,}\theta^{\dagger{i}}(\tau)(t^bt^a)^{ij}\bigl(\bar{\psi}_{\alpha}\Psi_{\alpha}^{j}(x)\bigr)-
f_0^{\ast}\bigl(\bar{\Psi}^{j}_{\alpha}(x)\psi_{\alpha}\bigr)(t^at^b)^{ji}\theta^{i}(\tau)\Bigr\}=0.
\end{split}
\]
\indent
Finally, varying with respect to $x^{\mu}$, we obtain the remaining equation of motion
\begin{equation}
m\ddot{x}_{\mu}(\tau)-gQ^a(\tau)F^{a}_{\mu\nu}(x)\dot{x}^{\nu}(\tau)
\label{eq:2i}
\end{equation}
\[
-\,\frac{g}{\sqrt{2}\,m}\Bigl\{\bar{\psi}_{\alpha}\Bigl(\theta^{\dagger{i}}(\tau)\overrightarrow{D}^{ij}_{\mu}(x)\Psi^{j}_{\alpha}(x)\Bigr)
+\Bigl(\bar{\Psi}^{j}_{\alpha}(x)\overleftarrow{D}^{\dagger{ji}}_{\mu}(x)\theta^i(\tau)\Bigr)\psi_{\alpha}\Bigr\}
\]
\[
+\,\frac{g}{\sqrt{2}\,m}\,Q^a(\tau)\Bigl\{\!f_{0\,}\bar{\psi}_{\alpha}\Bigl(\theta^{\dagger{i}}(\tau)(t^a)^{ij}\overrightarrow{D}^{jk}_{\mu}(x)\Psi^{k}_{\alpha}(x)\Bigr)
+f_0^{\ast}\Bigl(\bar{\Psi}^{k}_{\alpha}(x)\overleftarrow{D}^{\dagger{kj}}_{\mu}(x)(t^a)^{ji}\theta^i(\tau)\Bigr)\psi_{\alpha}\Bigr\}=0,
\]
where $F^{a}_{\mu\nu}=\partial_{\mu}A^a_{\nu}-\partial_{\nu}A^a_{\mu}-gf^{abc}A^b_{\mu}A^c_{\nu}$ is the strength tensor and
\[
\begin{split}
&\overrightarrow{D}^{jk}_{\mu}(x)=\delta^{jk}\overrightarrow{\partial}\!/\partial x^{\mu}+igA^{a}_{\mu}(x)(t^{a})^{jk}\,(\equiv{D}^{jk}_{\mu}(x)),\\
&\overleftarrow{D}^{\dagger{kj}}_{\mu}(x)=\delta^{kj}\overleftarrow{\partial}\!/\partial x^{\mu}-igA^{a}_{\mu}(x)(t^{a})^{kj}
\end{split}
\]
are the covariant derivations.
By adding the action of the kinetic energy of the background fields
\[
-\frac{1}{4}\int\!{d}^4{x}\,F^{a}_{\mu\nu}(x)F^{a\,\mu\nu}(x)+\,i\!\!\int\!{d}^4x\,\bar{\Psi}^{i}_{\alpha}(x)\gamma^{\mu}_{\alpha\beta}D^{ij}_{\mu}(x)\Psi^{j}_{\beta}(x)
\]
to the action (\ref{eq:2q}) and making the fields dynamical, one can obtain the Yang-Mills equation
\begin{equation}
D^{ab}_{\mu}(x)F^{b\,\mu\nu}(x)=j^{a\nu}(x),
\label{eq:2o}
\end{equation}
where ${D}^{ab}_{\mu}(x)=\delta^{ab}\partial/\partial x^{\mu}+igA^{c}_{\mu}(x)(T^{c})^{ab}$ is the covariant derivative in the adjoint representation,
\begin{equation}
j^{a\mu}(x)=g\!\!\int\!\dot{x}^{\mu}(\tau)\hspace{0.03cm}Q^{a}(\tau)\,\delta^{(4)}(x-x(\tau))\,d\tau+g\bar{\Psi}(x)\gamma^{\mu}t^{a}\Psi(x)
\label{eq:2p}
\end{equation}
is the color current, and the Dirac equation
\begin{equation}
i\gamma^{\mu}_{\alpha\beta}D^{ij}_{\mu}(x)\Psi^{j}_{\beta}(x)=\eta^{i}_{\alpha}(x),
\label{eq:2a}
\end{equation}
where on the right-hand side, the function $\eta^{i}_{\alpha}(x)$ (which in the subsequent discussion will be referred to as the {\it color source}) is
\begin{equation}
\eta^{i}_{\alpha}(x)=\frac{g}{\sqrt{2}\,m}\int\Bigl\{\psi_{\alpha\,}\theta^{i}(\tau) - \psi_{\alpha}f_0^{\ast}(\tau)\hspace{0.03cm}Q^a(\tau)(t^a)^{ij}\theta^j(\tau)\hspace{0.03cm}\!\Bigr\}\,
\delta^{(4)}(x-x(\tau))\,d\tau.
\label{eq:2s}
\end{equation}

\section{\bf Real Grassmann color charges $\vartheta^a$}
\setcounter{equation}{0}

As was shown in \cite{balachandran_1977, barducci_1977} instead of a set of the Grassmann color charges $\theta^{i}$ and $\theta^{\dagger{i}}$, belonging to the fundamental representation,
a set of the real Grassmann charges $\vartheta^a$, $a=1,\ldots,N_c^2-1$ belonging to the adjoint representation of the $SU(N_c)$ group, can also be introduced into consideration. Now instead of the Lagrangian (\ref{eq:1w}) we will have
\begin{equation}
L_{\vartheta}=-m\sqrt{\dot{x}^{\mu}\dot{x}_{\mu}}+\frac{i}{2}\,\vartheta^aD^{ab}\vartheta^b.
\label{eq:3q}
\end{equation}
Here, $D^{ab}=\delta^{ab}\partial/\partial\tau+ig\dot{x}^{\mu}A^{c}_{\mu}(T^{c})^{ab}$ and correspondingly the equations of motion are
\[
\begin{split}
&m\,\frac{d}{d\tau}\!\left(\frac{\dot{x}^{\mu}}{(\dot{x}^{\nu}\dot{x}_{\nu})^{1/2}}\right)=
g{\cal Q}^aF^{a}_{\mu\nu\,}\dot{x}^{\nu},\\
&\frac{d\vartheta^a(\tau)}{d\tau} + \,ig\dot{x}^{\mu}(\tau)A_{\mu}^{b}(x)(T^b)^{ac}\vartheta^c(\tau)=0.
\end{split}
\]
In this case instead of (\ref{eq:2e}) the commiting color charge is defined by the following expression:
\begin{equation}
{\cal Q}^a\equiv\frac{1}{2}\,\vartheta^b(T^a)^{bc}\vartheta^c.
\label{eq:3w}
\end{equation}
To distinguish this color charge from (\ref{eq:2e}), we have used a calligraphic capital letter ${\cal Q}$ instead of the usual one $Q$. It is easily to see that in the absence of the background fermion field these two descriptions of the color degree of freedom are completely equivalent\footnote{\,Thus in the expression for color current (\ref{eq:2p}) the Grassmann charges appear only in combinations (\ref{eq:2e}) or (\ref{eq:3w}). Furthermore, only these combinations enter into the generalized Newton equation and also in the equation describing spin dynamics \cite{balachandran_1977, barducci_1977}.}. This means that on the classical level of a description of the color dynamics of a color-charged particle in the absence of the background fermion field, we can not distinguish to which of the group representations this particle belongs.\\
\indent
However, this equivalence can be expected to disappear when we introduce a backgro\-und fermion field in the system. Under such conditions, for a fuller treatment of the dynamics of a classical color particle in external non-Abelian fields, it may be required to use of a set of the Grassmann charges belonging to different representations of the $SU(N_c)$ group. In this case, the simplest
extension of Lagrangian (\ref{eq:2w}) taking into account the fact just mentioned has the following form:
\[
L = L_{\theta} + L_{\vartheta},
\]
where $L_{\theta}$ is given by the expression (\ref{eq:2w}) and $L_{\vartheta}$ is defined by
\begin{equation}
L_{\vartheta}=\frac{i}{2}\,\vartheta^aD^{ab}\vartheta^b + \frac{e}{\sqrt{2}}\;g{\cal Q}^a\Bigl\{\tilde{f}_{0\,}\theta^{\dagger{i}}(t^{a})^{ij}\bigl(\bar\psi_{\alpha}\Psi^{j}_{\alpha}\bigr)
+\tilde{f}_0^{\ast}\bigl(\bar\Psi^{j}_{\alpha}\psi_{\alpha}\bigr)(t^{a})^{ji}\theta^{i}\Bigr\}.
\label{eq:3r}
\end{equation}
Now the equation of motion for the color charge $\vartheta^a$ is (in the gauge $e=1/m$)
\begin{equation}
\frac{d\vartheta^a(\tau)}{d\tau}
+ig\hspace{0.01cm}\dot{x}^{\mu}(\tau)A_{\mu}^{b}(x)(T^b)^{ac}\vartheta^b(\tau)
\label{eq:3t}
\end{equation}
\[
-\,\frac{ig}{\sqrt{2}\,m}\,(T^{b})^{ac}\vartheta^c(\tau)\Bigl\{\tilde{f}_{0\,}(\tau)\theta^{\dagger{i}}(\tau)(t^{b})^{ij}(\bar\psi_{\alpha}\Psi^{j}_{\alpha}(x))
+\tilde{f}_0^{\ast}(\tau)(\bar\Psi^{j}_{\alpha}(x)\psi_{\alpha})(t^{b})^{ji}\theta^{i}(\tau)\Bigr\}=0.
\]
The complex function $\tilde{f}_0$  is an arbitrary gauge and reparametrization invariant function, which generally speaking, does not coincide with a similar function $f_0$ in the Lagrangian (\ref{eq:2w}). From equation (\ref{eq:3t}), it is easy to obtain the equation of motion for the usual color charge ${\cal Q}^a$:
\begin{equation}
\frac{d{\cal Q}^a(\tau)}{d\tau} + ig\hspace{0.01cm}\dot{x}^{\mu}(\tau)A^b_{\mu}(x)(T^b)^{ac}{\cal Q}^c(\tau)
\label{eq:3tt}
\end{equation}
\[
-\,\frac{ig}{\sqrt{2}\,m}\,(T^b)^{ac}{\cal Q}^c(\tau)\Bigl\{\tilde{f}_{0\,}\theta^{\dagger{i}}(\tau)(t^{b})^{ij}
\bigl(\bar\psi_{\alpha}\Psi^{j}_{\alpha}(x)\bigr) +
\tilde{f}_0^{\ast}\bigl(\bar\Psi^{j}_{\alpha}(x)\psi_{\alpha}\bigr)(t^{b})^{ji}\theta^{i}(\tau)\Bigr\}=0.
\]
Besides, the additive Lagrangian (\ref{eq:3r}) results in appearing new terms in the equations of motion obtained in the previous section. Thus, in equation (\ref{eq:2t}) for the Grassmann charge
$\theta^i$ it is necessary to add the term
\begin{equation}
-\frac{ig}{\sqrt{2}\,m}\,\tilde{f}_{0\,}{\cal Q}^a(t^{a})^{ij}(\bar\psi_{\alpha}\Psi^{j}_{\alpha})
\label{eq:3y}
\end{equation}
and a proper term is added to the conjugate equation (\ref{eq:2y}). Furthermore, in equation (\ref{eq:2u}) for the classical commutating charge $Q^a$, a new contribution arises:
\begin{equation}
-\,\frac{ig}{\sqrt{2}\,m}\,{\cal Q}^b(\tau)\Bigl\{\tilde{f}_{0\,}\theta^{\dagger{i}}(\tau)(t^at^b)^{ij}\bigl(\bar{\psi}_{\alpha}\Psi_{\alpha}^{j}(x)\bigr)-
\tilde{f}_0^{\ast}\bigl(\bar{\Psi}^{j}_{\alpha}(x)\psi_{\alpha}\bigr)(t^bt^a)^{ji}\theta^{i}(\tau)\Bigr\}\equiv
\label{eq:3u}
\end{equation}
\[
\begin{split}
\equiv\;
+\,&\frac{ig}{2\sqrt{2}\,m}\;(T^b)^{ac}{\cal Q}^c(\tau)\!
\left\{\tilde{f}_{0\,}\theta^{\dagger{i}}(\tau)(t^b)^{ij}\!\left(\bar{\psi}_{\alpha}\Psi_{\alpha}^{j}(x)\right)+
\tilde{f}_0^{\ast}\left(\bar{\Psi}^{j}_{\alpha}(x)\psi_{\alpha}\right)\!(t^b)^{ji}\theta^{i}\right\}\\
-\,&\frac{ig}{2\sqrt{2}\,m}\;({\cal P}^b)^{ac}{\cal Q}^c(\tau)\!
\left\{\tilde{f}_{0\,}\theta^{\dagger{i}}(\tau)(t^b)^{ij}\!\left(\bar{\psi}_{\alpha}\Psi_{\alpha}^{j}(x)\right)-
\tilde{f}_0^{\ast}\left(\bar{\Psi}^{j}_{\alpha}(x)\psi_{\alpha}\right)\!(t^b)^{ji}\theta^{i}\right\}
\end{split}
\]
\[
-\,\frac{ig}{2\sqrt{2}\,m}\;\frac{1}{N_c}\,{\cal Q}^a(\tau)\!
\left\{\tilde{f}_{0\,}\theta^{\dagger{i}}(\tau)\!\left(\bar{\psi}_{\alpha}\Psi_{\alpha}^{i}(x)\right)-
\tilde{f}_0^{\ast}\left(\bar{\Psi}^{i}_{\alpha}(x)\psi_{\alpha}\right)\!\theta^{i}(\tau)\right\},
\]
where $({\cal P}^a)^{bc}\equiv d^{abc}$. We have to add also the term
\begin{equation}
\frac{g}{\sqrt{2}\,m}\,{\cal Q}^a(\tau)\Bigl\{\!f_{0\,}\bar{\psi}_{\alpha}\Bigl(\theta^{\dagger{i}}(\tau)(t^a)^{ij}\overrightarrow{D}^{jk}_{\mu}(x)\Psi^{k}_{\alpha}(x)\Bigr)
+f_0^{\ast}\Bigl(\bar{\Psi}^{k}_{\alpha}(x)\overleftarrow{D}^{\dagger{kj}}_{\mu}(x)(t^a)^{ji}\theta^i(\tau)\Bigr)\psi_{\alpha}\Bigr\}
\label{eq:3i}
\end{equation}
to the equation for the position $x^{\mu}$ of a color charge, (\ref{eq:2i}). Finally, we must add the terms
\begin{equation}
\begin{split}
g\!\!&\int\!\dot{x}^{\mu}(\tau)\hspace{0.03cm}{\cal Q}^{a}(\tau)\,\delta^{(4)}(x-x(\tau))\,{d}\tau,\\
-\frac{g}{\sqrt{2}\,m}&\int\!\psi_{\alpha}\tilde{f}_0^{\ast}(\tau)\hspace{0.03cm}{\cal Q}^a(\tau)(t^a)^{ij}\theta^j(\tau)\,\delta^{(4)}(x-x(\tau))\,{d}\tau
\end{split}
\label{eq:3o}
\end{equation}
to the right-hand side of the field equations, (\ref{eq:2o}) and (\ref{eq:2a}), correspon\-ding\-ly.

\section{\bf The most general gauge-invariant Lagrangian: the case of the linear dependence on the $\bar{\Psi}^i_{\alpha}$- and ${\Psi}^i_{\alpha}$-\hspace{0.03cm}fields}
\setcounter{equation}{0}

This section is devoted to analysis of the most general structure of a color particle Lagrangian which satisfies the requirement of gauge invariance. We will follow the arguments in Barducci
{\it et al} \cite{barducci_1977} closely. \\
\indent
In the general case, the desired Lagrangian is a function of the following variables:
\[
L=L(\dot{x}^{\mu};\theta^{\dagger{i}}\!, \theta^i\!, \dot{\theta}^{\dagger{i}}\!, \dot{\theta}^i\!, \vartheta^{a}\!,\dot{\vartheta}^{a};\bar{\psi}_{\alpha},\psi_{\alpha};
A_{\mu}^a,A_{\mu,\,\nu}^a,\bar{\Psi}^{i}_{\alpha},\Psi^{i}_{\alpha}).
\]
As was mentioned in Section 2, in the present work we do not consider a change of the spin state of the particle; therefore, the dependence of $L$ on $\dot{\bar{\psi}}_{\alpha}$ and $\dot{\psi}_{\alpha}$ is omitted. The total variation of the Lagrangian under the infinitesimal changes of all dynamical variables looks as follows:
\begin{equation}
\begin{split}
\delta{L}=\,&\frac{\overrightarrow{\partial}\!{L}}{\partial{\theta^{i}}}\,\delta{\theta^{i}}
+\delta{\theta^{\dagger{i}}}\frac{\overleftarrow{\partial}\!{L}}{\partial{\theta^{\dagger{i}}}}
+\frac{\overrightarrow{\partial}\!{L}}{\partial{\vartheta^{a}}}\,\delta{\vartheta^{a}}
+\frac{\overrightarrow{\partial}\!{L}}{\partial{\dot{\theta}^{i}}}\,\delta{\dot{\theta}^{i}}
+\delta{\dot{\theta}^{\dagger{i}}}\frac{\overleftarrow{\partial}\!{L}}{\partial{\dot{\theta}^{\dagger{i}}}}
+\frac{\overrightarrow{\partial}\!{L}}{\partial{\dot{\vartheta}^{a}}}\,\delta{\dot{\vartheta}^{a}}\\
&+\frac{\overrightarrow{\partial}\!{L}}{\partial{\Psi^{i}_{\alpha}}}\,\delta{\Psi^{i}_{\alpha}}
+\delta{\bar{\Psi}^{i}_{\alpha}}\frac{\overleftarrow{\partial}\!{L}}{\partial{\bar{\Psi}^{i}_{\alpha}}}
+\frac{{\partial}{L}}{\partial{A_{\mu}^a}}\,\delta{A^a_{\mu}}
+\frac{{\partial}{L}}{\partial{A_{\mu,\,\nu}^a}}\,\delta{A^a_{\mu,\,\nu}}
=0.
\end{split}
\label{eq:4q}
\end{equation}
Here, the right (left) arrow above the partial derivatives with respect to corresponding Grassmann variables indicates that the derivative acts from the right (left) on the Lagrangian.
The variation of the Lagrangian associated with the gauge transformation of the background boson field $A^{a}_{\mu}(x)$ has been studied in detail in \cite{barducci_1977}. In particular, it has been shown that the gauge potential (and its derivatives) enters into the Lagrangian in question by a gauge covariant manner through the covariant field tensor $F^{a}_{\mu\nu}$ and in covariant combinations with variables $\dot\theta^i,\,\dot\theta^{\dagger i}$ and $\dot\vartheta^a$ of the form
\[
\dot\theta^i+ig\dot{x}^{\mu\!}A^{b}_{\mu}(t^b)^{ij}\theta^j,{\quad}\dot\vartheta^a+ig\dot{x}^{\mu\!}A^{b}_{\mu}(T^b)^{ac}\vartheta^c.
\]
At present, we will confine our attention to the variation of the Lagrangian with respect to the gauge transformation of the background fermion fields. Therefore, in what follows we will not explicitly write down the dependence of the Lagrangian on the Grassmann variables $\dot\theta^i$, $\dot\theta^{\dagger i}$, $\dot\vartheta^a$ and the gauge field $A^{a}_{\mu}$. For simplicity, in most cases we will not also write down in an explicit form the dependence of $L$ on the spinors $\bar{\psi}_{\alpha}$ and $\psi_{\alpha}$.\\
\indent
Substituting infinitesimal gauge transformations (\ref{eq:2r}) into equation (\ref{eq:4q}) for variations of dynamical variables, we obtain the following condition of gauge invariance of the  Lagrangian:
\begin{equation}
\delta_{\Lambda}{L}=
\label{eq:4w}
\end{equation}
\[
=ig\Lambda^{a}\!\left(\frac{\overrightarrow{\partial}\!{L}}{\partial{\theta^{i}}}(t^{a})^{ij}\theta^{j}
-\theta^{\dagger{j}}(t^a)^{ji}\frac{\overleftarrow{\partial}\!{L}}{\partial{\theta^{\dagger{i}}}}
-\frac{\overrightarrow{\partial}\!{L}}{\partial{\vartheta^{b}}}(T^b)^{ac}\vartheta^{c}
+\frac{\overrightarrow{\partial}\!{L}}{\partial{\Psi^{i}_{\alpha}}}(t^a)^{ij}\Psi^{j}_{\alpha}
-\bar{\Psi}^{j}_{\alpha}(t^a)^{ji}\frac{\overleftarrow{\partial}\!{L}}{\partial{\bar{\Psi}^{i}_{\alpha}}}\right)=0.
\]
\indent
In this section, we restrict ourselves to the case of the linear dependence of the Lagrangian on the background fermion fields $\bar{\Psi}^{i}_{\alpha}$ and $\Psi^{i}_{\alpha}$. Let us first consider the most simple case when these Grassmann-valued fields enter into the Lagrangian only in the following combinations: $(\bar{\Psi}^{i}_{\alpha}\psi_{\alpha})$ and $(\bar{\psi}_{\alpha}\Psi^{i}_{\alpha})$, i.e. we set
\[
L_{\Psi,\bar{\Psi}}={\cal L}^{\dagger i}(\theta^{\dagger}\!,\theta,\vartheta)(\bar{\psi}_{\alpha}\Psi_{\alpha}^{i})+
(\bar{\Psi}_{\alpha}^{i}\psi_{\alpha}){\cal L}^{i}(\theta^{\dagger}\!,\theta,\vartheta).
\]
If we substitute the previous expression into equation (\ref{eq:4w}), then the requirement of gauge invariance reduces to two equations for the functions ${\cal L}^i$ and ${\cal L}^{\dagger{i}}$:
\begin{equation}
\begin{split}
&-\frac{\overrightarrow{\partial}\!{\cal L}^{\dagger i}}{\partial{\theta}^{j}}\,(t^a)^{jk}\theta^k
+\theta^{\dagger{k}}(t^a)^{kj}\,\frac{\overleftarrow{\partial}\!{\cal L}^{\dagger i}}{\partial{\theta}^{\dagger{j}}}
-{\cal L}^{\dagger j}(t^a)^{ji}
+\frac{\overrightarrow{\partial}\!{\cal L}^{\dagger i}}{\partial{\vartheta}^{b}}\,(T^b)^{ac}\vartheta^c=0,\\
&-\frac{\overrightarrow{\partial}\!{\cal L}^{i}}{\partial{\theta}^{j}}\,(t^a)^{jk}\theta^k
+\theta^{\dagger{k}}(t^a)^{kj}\,\frac{\overleftarrow{\partial}\!{\cal L}^{i}}{\partial{\theta}^{\dagger{j}}}
+(t^a)^{ij}{\cal L}^{j}
+\frac{\overrightarrow{\partial}\!{\cal L}^{i}}{\partial{\vartheta}^{b}}\,(T^b)^{ac}\vartheta^c=0.
\label{eq:4e}
\end{split}
\end{equation}
It is easy to verify that the following combinations of color charges:
\begin{equation}
\theta^{\dagger{i}},\; Q^{a}\theta^{\dagger{j}}(t^a)^{ji}{\,\rm and\;} {\cal Q}^a\theta^{\dagger{j}}(t^{a})^{ji}
\label{eq:4r}
\end{equation}
satisfy the first equation in (\ref{eq:4e}), and correspondingly, conjugate combinations satisfy the second equation. It is these combinations that are presented in the expressions of the
Lagrangians (\ref{eq:2w}) and (\ref{eq:3r}). We have not found any other more complicated in color structure solutions containing greater number of Grassmann charges and Hermitian generators  $t^a$ and $T^a$.\\
\indent
However, a more subtle analysis of the general conditions of gauge invariance (\ref{eq:4e}) has shown that the color charge combinations (\ref{eq:4r}) may enter into the Lagrangian with some arbitrary scalar function $f_0$ as a multiplier. This function depends on gauge-invariant combinations of Grassmann charges:
\[
f_0=f_0(\theta^{\dagger}\!,\theta,\vartheta)\equiv f_0(\theta^{\dagger{i}}\theta^{i}\!,Q^{a}Q^{a}\!,{\cal Q}^a{\cal Q}^a).
\]
In the case of the absence of a background fermion field, each of these colorless quadratic combinations of charges is conserved by virtue of the relevant equation of motion and therefore the function $f_0$ merely represents itself a numerical factor. From the other hand, if there exists a background fermion field, then the quadratic combinations can be rather complicated functions of $\tau$, so that $f_0\,{\neq}\;$const.
It is worthy of special emphasis that the function $f_0$ is broadly speaking different for each of three solutions (\ref{eq:4r}) and its explicit form is not fixed by the requirements specified in the Introduction. In the general case, the function represents a finite polynomial of its independent variables with arbitrary coefficients, whose an explicit form should be determined from some additional considerations\footnote{\,For example, in \cite{markov_NPA_2007} we have shown that to obtain gauge-invariant expressions for matrix elements of some scattering processes of soft  quark-gluon plasma excitations off hard thermal particles, an arbitrary function $f_0$ in the fourth term in the Lagrangian (\ref{eq:2w}) must be in exactly equal to $-1$. In Section 6 of the present work, another consistency condition will fix the function $f_0$ in the last term in (\ref{eq:2w}).}. Besides, for the group $SU(N_c)$ with $N_c\geqslant 3$, this function can be also dependent on more complicated gauge invariants of the form:
\[
d^{abc}Q^aQ^bQ^c {\;\;\rm and\;\;} d^{abc}{\cal Q}^a{\cal Q}^b{\cal Q}^c,
\]
where $d^{abc}$ is the totally symmetric structure constant of the group.\\
\indent
We now proceed to a discussion of the general structure of the Lagrangian with the linear dependence on external non-Abelian fermion fields.
First, let us write down the most general spinor structure of the given Lagrangian:
\begin{equation}
L_{\Psi,\bar{\Psi}}={\cal L}^{\dagger i}_A(\dot{x},\theta^{\dagger}\!,\theta,\vartheta,F_{\mu\nu})(\bar{\psi}_{\alpha}\Gamma^A_{\alpha\beta}\Psi_{\beta}^{i})+
(\bar{\Psi}_{\beta}^{i}\Gamma^A_{\beta\alpha}\psi_{\alpha})\,{\cal L}^{i}_A(\dot{x},\theta^{\dagger}\!,\theta,\vartheta,F_{\mu\nu}),
\label{eq:4t}
\end{equation}
where $\Gamma^A$ is any one of these 16 independent generators of the Clifford algebra:
\[
I,\; \gamma^5\!,\; \gamma^{\mu}\!,\; i \gamma^{\mu} \gamma^5\!,\; \sigma^{\mu \nu}.
\]
It is just impossible to make up vector or tensor quantities only from the Grassmann-valued charges. Therefore, among arguments of  the functions ${\cal L}_{A}^{\dagger i}$ and ${\cal L}_{A}^i$, we explicitly specify those vector and tensor functions, which we have at our disposal.\\
\indent
At the beginning, we deal with the case of the $\gamma^{\mu}$ matrices instead of $\Gamma^{A}$. The simplest structure of ${\cal L}_{\mu}^{\dagger i}$ (or ${\cal L}_{\mu}^i$) represents a perfect factorization with respect to color and vector indices, i.e.
\begin{equation}
{\cal L}_{\mu}^{\dagger i} = {\cal L}^{\dagger i} (\theta^{\dagger}\!,\theta,\vartheta)\,{\cal L}_{\mu}^{\ast}(\dot{x}, \bar{\psi}, \psi)
\label{eq:4y}
\end{equation}
and a similar equation is true for the ${\cal L}_{\mu}^i$ function. As ${\cal L}^{\dagger i}$, any one of the combinations of color charges in a set (\ref{eq:4r}), can be taken, and as
${\cal L}_{\mu}^{\ast}$ the function $\dot{x}_{\mu}$ can be taken. Thus, for example, for the first two color structures in (\ref{eq:4r}) we have
the following gauge (and reparametrization) invariant contributions which should be added to the required Lagrangian:
\[
g\dot{x}_{\mu}\bigl\{f_{0\,}\theta^{\dagger{i}}(\bar\psi\gamma^{\mu}\Psi^{i}) + f^{\ast}_0(\bar\Psi^{i}\gamma^{\mu}\psi)\theta^{i}\bigr\},
\]
\[
g\dot{x}_{\mu}Q^a\bigl\{f_{0\,}\theta^{\dagger{i}}(t^{a})^{ij}(\bar\psi\gamma^{\mu}\Psi^{j}) + f^{\ast}_0(\bar\Psi^{j}\gamma^{\mu}\psi)(t^{a})^{ji}\theta^{i}\bigr\}.
\]
Besides, there exist the gauge and reparametrization invariant contributions in $L_{\Psi,\bar{\Psi}}$ at $\Gamma^A \equiv \gamma^{\mu}$ for which there is no factorization of the type (\ref{eq:4y}).
However, in this case it should be invoked the strength tensor of background gauge field. The following expression
\[
g\dot{x}^{\mu}F^a_{\mu\nu}\bigl\{f_{0\,}\theta^{\dagger{i}}(t^{a})^{ij}(\bar\psi\gamma^{\nu}\Psi^{j}) + f^{\ast}_0(\bar\Psi^{j}\gamma^{\nu}\psi)(t^{a})^{ji}\theta^{i}\bigr\}
\]
provides an example of such type of contribution.\\
\indent
Furthermore, the necessity to invoke the field strength arises also when the $\Gamma^A$ is taken in the form $\sigma^{\mu \nu}$. Here, we will have the gauge-invariant contributions of the following form:
\[
\frac{g}{\sqrt{2}\,m}\,F^a_{\mu\nu}\Bigl\{f_{0\,}\theta^{\dagger{i}}(t^{a})^{ij}(\bar\psi\sigma^{\mu\nu}\Psi^{j}) + f^{\ast}_0(\bar\Psi^{j}\sigma^{\mu\nu}\psi)(t^{a})^{ji}\theta^{i}\Bigr\},
\]
\[
\frac{g}{\sqrt{2}\,m}\,Q^aF^a_{\mu\nu}\Bigl\{f_{0\,}\theta^{\dagger{i}}(\bar\psi\sigma^{\mu\nu}\Psi^{i}) + f^{\ast}_0(\bar\Psi^{i}\sigma^{\mu\nu}\psi)\theta^{i}\Bigr\},
\]
and also the contributions from the higher order terms in powers of $F_{\mu \nu}^a$. All these terms represent examples of spin-color interaction.\\
\indent
As regards the terms with the $\gamma^5$ matrix in a set $\Gamma^A$, pseudoscalar and pseudovector functions of the type $(\bar{\psi}\gamma^5\Psi^i),\, i(\bar{\psi}\gamma^{\mu}\gamma^5\Psi^i),\,\dots$ may be `canceled' only by factors of the form $(\bar{\psi} \gamma^5 \psi)$, $i(\bar{\psi}\gamma_{\mu}\gamma^5 \psi),\,\ldots\;$.
Though most likely such types of contributions should be rejected as unrelated to real interaction processes in the problem under consideration.\\
\indent
%%%%%%%%%%%%%%%%%%%%%%%%%%%%%%%%%%%%%%%%%%%%%%%%%%%%%%%%%%%%%%%%%%%%%%%%%%%%%%%%%%%%%%%%%%%%%%%%%%%%%%%%%%%%%%%%%%%%%%%%%%%%%%%%%%%%%%%%%%%%%%%%%%%%%%%%%%%%%%%%%%%%%%%%%%%%%%%%%%%%%%%%%%%%%%%%%%%%%%%
One can rise a question about the dependence of the interaction Lagrangian on derivatives of background fermionic fields, i.e. on the functions of the type
$\partial_{\mu}\bar{\Psi}^i_{\alpha}(\equiv \bar{\Psi}^i_{\alpha,\,\mu})$ and $\partial_{\mu}\Psi^i_{\alpha}(\equiv{\Psi}^i_{\alpha,\,\mu})$. We again restrict our consideration to the case of the linear dependence and when these functions appear into the Lagrangian only in the simplest combinations: $(\partial_{\mu}\bar{\Psi}^i_{\alpha}\psi_{\alpha})$ and $(\bar{\psi}_{\alpha}\partial_{\mu}\Psi^i_{\alpha})$, i.e. we set
\begin{equation}
L_{\partial\Psi,\,\partial\bar{\Psi}}={\cal L}^{\dagger{i}}_{\mu}(\dot{x},\theta^{\dagger}\!,\theta,\vartheta)
(\bar{\psi}_{\alpha}\partial^{\mu}\Psi_{\alpha}^{i}(x)) +
(\partial^{\mu}\bar{\Psi}_{\alpha}^{i}(x)\psi_{\alpha})
{\cal L}^{i}_{\mu}(\dot{x},\theta^{\dagger}\!,\theta,\vartheta).
\label{eq:4u}
\end{equation}
It is clear that the derivatives of the $\Psi$-fields can enter into the Lagrangian in question only in the form of the covariant derivatives; therefore, instead of (\ref{eq:4u}) we can write down at once
\begin{equation}
L_{\overrightarrow{D}\Psi,\bar{\Psi}\overleftarrow{D}^{\dagger}}={\cal L}^{\dagger{i}{\mu}}(\dot{x}, \theta^{\dagger}\!,\theta,\vartheta)
\bigl(\bar{\psi}_{\alpha}\overrightarrow{D}^{ij}_{\mu}(x)\Psi_{\alpha}^{j}(x)\bigr) +
\bigl(\bar{\Psi}_{\alpha}^{j}(x)\overleftarrow{D}^{\dagger{ji}}_{\mu}(x)\psi_{\alpha}\bigr)
{\cal L}^{i\mu}(\dot{x},\theta^{\dagger}\!,\theta,\vartheta).
\label{eq:4i}
\end{equation}
The dependence of the Lagrangian on the $\Psi$-field derivatives leads to the fact that in the total variation (\ref{eq:4q}) the terms of the following form
\[
\frac{\overrightarrow{\partial}\!{L}}{\partial{\Psi^{i}_{\alpha,\,\mu}}}(t^a)^{ij}\delta\Psi^{j}_{\alpha,\,\mu}
-\,\delta\bar{\Psi}^{j}_{\alpha,\,\mu}(t^a)^{ji}\frac{\overleftarrow{\partial}\!{L}}{\partial{\bar{\Psi}^{i}_{\alpha,\,\mu}}}
\]
must be added, where in a particular case of variations induced by the infinitesimal gauge transformation, we have
\[
\begin{split}
\delta\Psi^{i}_{\alpha,\,\mu} &= ig(\partial_{\mu}\Lambda^{a})(t^{a})^{ij}\Psi^{j}_{\alpha} +\, ig\Lambda^{a}(t^{a})^{ij}(\partial_{\mu}\Psi^{j}_{\alpha}),\\
\delta\bar{\Psi}^{i}_{\alpha,\,\mu} &= -ig(\partial_{\mu}\Lambda^{a})\bar{\Psi}^{j}_{\alpha}(t^{a})^{ji} - ig\Lambda^{a}(\partial_{\mu}\bar{\Psi}^{j}_{\alpha})(t^{a})^{ji}.
\end{split}
\]
These transformations should be added to the list (\ref{eq:2r}). It is easy to convince ourselves that the basic requirement of gauge independence of the Lagrangian (\ref{eq:4i}) is reduced to
the fulfilment of equations (\ref{eq:4e}), where we need to substitute ${\cal L}^{\dagger i}_{\mu}$ and ${\cal L}^i_{\mu}$ for ${\cal L}^{\dagger i}$ and ${\cal L}^{i}$. The vector index
appears in these equations in the parametrical manner and therefore without loss of generality, these functions can be taken in the factored form (\ref{eq:4y}). Following the same line of reasoning
given after equation (\ref{eq:4y}), we conclude that there exist only three independent gauge-invariant Lagrangians of the form (\ref{eq:4i}), namely
\[
\begin{split}
\dot{x}^{\mu}\Bigl\{\!f_0\,\bar{\psi}_{\alpha\,}\!\Bigl(\theta^{\dagger{i}}\overrightarrow{D}^{ij}_{\mu}(x)\Psi^{j}_{\alpha}(x)\Bigr)\!
&+f^{\ast}_0\Bigl(\bar{\Psi}^{j}_{\alpha}(x)\overleftarrow{D}^{\dagger{ji}}_{\mu}(x)\theta^i\Bigr)\psi_{\alpha}\!\Bigr\},\\
\dot{x}^{\mu}Q^a\Bigl\{\!f_0\,\bar{\psi}_{\alpha\,}\!\Bigl(\theta^{\dagger{i}}(t^a)^{ij}\overrightarrow{D}^{jk}_{\mu}(x)\Psi^{k}_{\alpha}(x)\Bigr)\!
&+f^{\ast}_0\Bigl(\bar{\Psi}^{k}_{\alpha}(x)\overleftarrow{D}^{\dagger{kj}}_{\mu}(x)(t^a)^{ji}\theta^i\Bigr)\psi_{\alpha}\!\Bigr\}
\end{split}
\]
and the last expression with the replacement $Q^a \rightarrow {\cal Q}^a$. These Lagrangians satisfy all requirements listed in the Introduction. Formally, we have to add them in the total interaction Lagrangian.\\
\indent
Finally, instead of (\ref{eq:4i}) we can consider a more general expression of the form
\begin{equation}
\begin{split}
L_{\overrightarrow{D}\Psi,\bar{\Psi}\overleftarrow{D}^{\dagger}}=\;
&{\cal L}^{\dagger{i}{\mu}}_A(\dot{x},\theta^{\dagger}\!,\theta,\vartheta,F_{\nu\lambda})
\bigl(\bar{\psi}_{\alpha}\Gamma^A_{\alpha\beta}\overrightarrow{D}^{ij}_{\mu}(x)\Psi_{\beta}^{j}(x)\bigr)\,+\\
+\,&\bigl(\bar{\Psi}_{\beta}^{j}(x)\overleftarrow{D}^{\dagger{ji}}_{\mu}(x)\Gamma^A_{\beta\alpha}\psi_{\alpha}\bigr)
{\cal L}^{i\mu}_A(\dot{x},\theta^{\dagger}\!,\theta,\vartheta,F_{\nu\lambda})
\end{split}
\label{eq:4o}
\end{equation}
in an exact analogy as was done for (\ref{eq:4t}). As in the above case (\ref{eq:4t}), the specific choice of $\Gamma^A$ ($\gamma^{\mu}$ or $\sigma^{\mu\nu}$) inevitably results in the necessity of introducing into consideration the strength tensor of background gauge field. This circumstance has been depicted in the notation of the
coefficient functions in (\ref{eq:4o}). In this case the number of independent contributions of the (\ref{eq:4o}) type becomes unlimited.

\section{\bf The most general gauge-invariant Lagrangian: the case of the quadratic dependence on the $\bar{\Psi}^i_{\alpha}$- and $\Psi^i_{\alpha}$-\hspace{0.03cm}fields}
\setcounter{equation}{0}

Let us analyze the case of the quadratic dependence of the Lagrangian on background fermionic field. To be specific, we consider in detail the most interesting dependence of the type
\begin{equation}
L_{\bar{\Psi}\Psi}=\bar{\Psi}^i_{\alpha\,}{\cal L}^{ij}_{\alpha\beta}(\theta^{\dagger}\!,\theta,\vartheta)\Psi^j_{\beta}.
\label{eq:5q}
\end{equation}
Substituting the above expression into the general condition of gauge invariance (\ref{eq:4w}), we obtain the following equation for the function ${\cal L}^{ij}_{\alpha\beta}$:
\begin{equation}
\frac{\overrightarrow{\partial}\!{\cal L}^{ij}_{\alpha\beta}}{\partial{\theta}^{k}}\,(t^a)^{ks}\theta^s
-\theta^{\dagger{s}}(t^a)^{sk}\,\frac{\overleftarrow{\partial}\!{\cal L}^{ij}_{\alpha\beta}}{\partial{\theta}^{\dagger{k}}}
-\frac{\overrightarrow{\partial}\!{\cal L}^{ij}_{\alpha\beta}}{\partial{\vartheta}^{b}}\,(T^b)^{ac}\vartheta^c+
(t^a)^{ik}{\cal L}^{kj}_{\alpha\beta} - {\cal L}^{ik}_{\alpha\beta}(t^a)^{kj} = 0.
\label{eq:5w}
\end{equation}
Here, the spinor indices appear in the parametric manner. By virtue of this fact, the spinor dependence of the function ${\cal L}^{ij}_{\alpha\beta}$ should be determined by some additional considerations\footnote{\,The only restriction here can arise if instead of the requirement of reality of the action we demand the fulfilment of a little more hard condition: reality of the Lagrangian (\ref{eq:5q}). Then in terms of the ${\cal L}^{ij}_{\alpha\beta}$ function this requirement leads to an equality
\[
{\cal L}^{ij}_{\alpha\beta} = \gamma^0_{\alpha \alpha^{\prime}} ({\cal L}^{ij}_{\alpha^{\prime}\!\beta^{\prime}})^{\ast}
\gamma^0_{\beta^{\prime}\beta}.
\]
For the decomposition (\ref{eq:5e}), i.e. for ${\cal L} = {\cal L}_{\rm spinor} \otimes {\cal L}_{\rm color}$, the last equality falls into two independent ones
\[
{\cal L}_{\rm spinor} = \gamma^0 {\cal L}^{\dagger}_{\rm spinor} \gamma_0,\quad {\cal L}_{\rm color} = {\cal L}^{\dagger}_{\rm color}.
\]
Here, $\otimes$ is a sign of the tensor product.
}.
In definition (\ref{eq:5q}) we assume for the moment that the ${\cal L}^{ij}_{\alpha\beta}$ is independent of the background bosonic field (but it is dependent on the spinors
$\bar{\psi}_{\alpha}$ and $\psi_{\alpha}$, see the previous section). For this reason, without loss of generality one can set
\begin{equation}
{\cal L}^{ij}_{\alpha\beta}\equiv{\cal L}^{ij}(\theta^{\dagger}\!,\theta,\vartheta)\,{\cal L}_{\alpha\beta}(\bar{\psi},\psi).
\label{eq:5e}
\end{equation}
As the ${\cal L}_{\alpha \beta}$ function we may choose any one of spinor structures of the form
\begin{equation}
{\cal L}_{\alpha\beta}\; \sim\; \delta_{\alpha\beta},\;
\psi_{\alpha}\bar{\psi}_{\beta},\;
(\gamma^5\psi)_{\alpha}(\bar{\psi}\gamma^5)_{\beta},\;
(\gamma^{\mu}\psi)_{\alpha}(\bar{\psi}\gamma_{\mu})_{\beta},\;
(\gamma^{\mu}\gamma^5\psi)_{\alpha}(\bar{\psi}\gamma^5\gamma_{\mu})_{\beta},\;
(\sigma^{\mu\nu}\psi)_{\alpha}(\bar{\psi}\sigma_{\mu\nu})_{\beta}.
\label{eq:5r}
\end{equation}
\indent
What can we take as the function ${\cal L}^{ij\,}$? The general analysis of equation (\ref{eq:5w}) has shown that there exists a considerable amount of independent color structures satisfying the requirement of gauge invariance, namely
\[
{\cal L}^{ij}\; \sim\;\;
\delta^{ij},\quad
\theta^i\theta^{\dagger j},\quad
(t^a\theta)^i(\theta^{\dagger} t^a)^{j},\quad
Q^a(t^a)^{ij},\quad
{\cal Q}^a(t^a)^{ij},
\]
\[
\vartheta^a(t^a\theta)^i(\theta^{\dagger} t^b)^{j}\vartheta^b,\quad
Q^a(t^a\theta)^i(\theta^{\dagger} t^b)^{j}Q^b,\quad
{\cal Q}^a(t^a\theta)^i(\theta^{\dagger} t^b)^{j}{\cal Q}^b,
\]
\[
\bigl\{Q^a(t^a\theta)^i(\theta^{\dagger} t^b)^{j}{\cal Q}^b+\,
{\cal Q}^a(t^a\theta)^i(\theta^{\dagger} t^b)^{j}Q^b\bigr\},
\]
\begin{equation}
if^{abc}Q^a(t^b\theta)^i(\theta^{\dagger} t^c)^{j},\quad
if^{abc}{\cal Q}^a(t^b\theta)^i(\theta^{\dagger} t^c)^{j},
\label{eq:5t}
\end{equation}
\[
\bigl[\,f^{abc}Q^b(t^c\theta)^i\bigr]\!
\bigl[\,f^{ade}Q^d(\theta^{\dagger} t^e)^{j}\bigl],\quad
\bigl[\,f^{abc}{\cal Q}^b(t^c\theta)^i\bigr]\!
\bigl[\,f^{ade}{\cal Q}^d(\theta^{\dagger} t^e)^{j}\bigl],
\]
\[
\Bigl\{\!\bigl[\,f^{abc}Q^b(t^c\theta)^i\bigr]\!
\bigl[\,f^{ade}{\cal Q}^d(\theta^{\dagger} t^e)^{j}\bigl]\,+\,
\bigl[\,f^{abc}{\cal Q}^b(t^c\theta)^i\bigr]\!
\bigl[\,f^{ade}Q^d(\theta^{\dagger} t^e)^{j}\bigl]\Bigr\},
\]
\[
\bigl[\,f^{abc}Q^a{\cal Q}^b(t^c\theta)^i\bigr]\!
\bigl[\,f^{def}(\theta^{\dagger} t^d)^{j}Q^e{\cal Q}^f\bigl],
\]
and so on. On substituting structures (\ref{eq:5t}) into equation (\ref{eq:5w}) the latter is reduced to either the identity or the relation for the totally antisym\-metric structure constants
\[
f^{adc}f^{bce}+f^{abc}f^{cde} = f^{bdc}f^{ace}.
\]
Hence in deciding on the Lagrangian in the form (\ref{eq:5q}) taking into account (\ref{eq:5r}) and (\ref{eq:5t}), we have considerably more (but nevertheless a finite number) gauge-invariant terms of interaction as compared with the sum of two Lagrangians (\ref{eq:2w}) and (\ref{eq:3r}). To choose the terms that are relevant to real dynamics of our physical system, one has to once again invoke other physical considerations. Let us note one remarkable feature of the second-fifth structures in (\ref{eq:5t}). These color structures depend on Grassmann $\theta$- and $\vartheta$-charges in the quadratic manner as well as the second terms in the Lagrangians (\ref{eq:2w}) and (\ref{eq:3r}). This circumstance results in an important modification of terms in the equations of motion (\ref{eq:2t}), (\ref{eq:2y}) and (\ref{eq:3t}) that are linear in the Grassmann charges. To be specific, let us consider the second spin structure in (\ref{eq:5r}), the third and fourth color ones in (\ref{eq:5t}). For this case instead of two terms in the first line of equation (\ref{eq:2t}), we have
\[
\frac{d\theta^i(\tau)}{d\tau}
\,+\,ig\dot{x}^{\mu\!}(\tau)A_{\mu}^{a}(x)(t^a)^{ij}\theta^j(\tau)-ig^2\bigl[(\bar\Psi^{k}_{\alpha}(x)\psi_{\alpha})(t^{a})^{ks}
(\bar\psi_{\beta}\Psi^{s}_{\beta}(x))\bigr](t^{a})^{ij}
\theta^{j}\!(\tau)
\]
\[
-\;ig^2\bigl[t^a(\bar{\psi}_{\beta}\Psi_{\beta}(x))\bigr]^i\bigl[(\bar{\Psi}_{\alpha}(x)\psi_{\alpha})t^a\bigr]^j\theta^j(\tau)
\,+\,\ldots\, = 0.
\]
New third and fourth terms in the above equation may be interpreted in terms of a modification (or extension) of the standard definition of the evolution operator, namely,
we have to consider the {\it extended} evolution operator
\[
{\cal U}(\tau,\tau_0)={\rm T}\exp\Biggl\{-ig\!\!\int\limits_{\tau_0}^{\tau}
\Bigl(\dot{x}^{\mu}({\tau}^{\prime})A^a_{\mu}(x({\tau}^{\prime}))\,t^a\,-
\hspace{0.4cm}
\]
\[
\hspace{0.8cm}
-\,g\bigl[(\bar\Psi_{\alpha}(x(\tau^{\prime}))\psi_{\alpha})\,t^{a} (\bar\psi_{\beta}\Psi_{\beta}(x(\tau^{\prime})))\bigr]t^{a}
-g\bigl[t^a(\bar\psi_{\beta}\Psi_{\beta}(x(\tau^{\prime})))\bigr]\!\otimes\!
\bigl[(\bar\Psi_{\alpha}(x(\tau^{\prime}))\psi_{\alpha})t^{a}\bigr]
\Bigr) d{\tau}^{\prime}\!\Biggr\}
\]
instead of
\begin{equation}
U(\tau,\tau_0)={\rm T}\exp\Biggl\{-ig\!\!\int\limits_{\tau_0}^{\tau}
\!\dot{x}^{\mu}({\tau}^{\prime})A^a_{\mu}(x({\tau}^{\prime}))\,t^a d{\tau}^{\prime}\!\Biggr\}.
\label{eq:5y}
\end{equation}
The possibility of appearance of such an extended evolution operator has been discussed in the papers \cite{markov_NPA_2007, markov_IJMPA_2010} in studing concrete physical scattering processes occurring in the quark-gluon plasma. The evolution operator ${\cal U}(t, t_0)$ takes into account the effect of rotation of a color charge vector in the internal color space induced by interaction with both gauge and fermion background fields.\\
\indent
If at this point we introduce into consideration the background bosonic field, then the choice of the spinor structures ${\cal L}_{\alpha\beta}$ becomes to a great extent richer and more varied. Thus, for example, to the structures (\ref{eq:5r}) one can add the terms of the type (see the last footnote)
\[
Q^{a\!} F^a_{\mu\nu}\bigl[(\sigma^{\mu\nu}\psi)_{\alpha}\bar{\psi}_{\beta}\,+\,
\psi_{\alpha}(\bar{\psi}\sigma^{\mu\nu})_{\beta}\bigr],\quad
Q^{a\!} F^a_{\mu\nu}(\sigma^{\mu\nu}\psi)_{\alpha}(\bar{\psi}\sigma^{\lambda\sigma})_{\beta}F^b_{\lambda\sigma}Q^b,
\]
\[
{\cal Q}^{a\!} F^a_{\mu\nu}(\sigma^{\mu\nu}\psi)_{\alpha}(\bar{\psi}\sigma^{\lambda\sigma})_{\beta}F^b_{\lambda\sigma}{\cal Q}^b,\quad
\vartheta^{a\!} F^a_{\mu\nu}(\sigma^{\mu\nu}\psi)_{\alpha}(\bar{\psi}\sigma^{\lambda\sigma})_{\beta}F^b_{\lambda\sigma}\vartheta^b,\,\ldots
\]
and the higher-order terms with respect to the strength tensor $F^a_{\mu\nu}$. Unlike (\ref{eq:5r}) this set includes already an infinite number of terms. Furthermore, the introduction of external  gauge field into the system enables us to consider another more interesting type of factorization of the ${\cal L}^{ij}_{\alpha\beta}$ function, namely, the factorization with respect to pairs of indices containing both color and spinor index, i.e.
\[
{\cal L}^{ij}_{\alpha\beta}\equiv
{\cal L}^{i}_{\alpha}(F_{\mu\nu},\psi,\theta)\,
{\cal L}^{j}_{\beta}(F_{\lambda\sigma},\bar{\psi},\theta^{\dagger}).
\]
The following two expressions can be considered as examples of such a factorization:
\[
\bigl[F^a_{\mu\nu}(\sigma^{\mu\nu}\psi)_{\alpha}(t^a\theta)^i\bigr]\!
\bigl[(\bar{\psi}\sigma^{\lambda\sigma})_{\beta}(\theta^{\dagger} t^b)^jF^b_{\lambda\sigma}\bigr],
\]
\[
Q^a\!f^{abc}\bigl[F^b_{\mu\nu}(\sigma^{\mu\nu}\psi)_{\alpha}(t^c\theta)^i\bigr]\!
\bigl[(\bar{\psi}\sigma^{\lambda\sigma})_{\beta}(\theta^{\dagger} t^e)^jF^f_{\lambda\sigma}\bigr]f^{efd}Q^d.
\]
Finally, the expression below represents the simplest example when we do not observe any factorization of the ${\cal L}^{ij}_{\alpha\beta}$ function
\[
\bigl[f^{abe}F^a_{\mu\nu}(\sigma^{\mu\nu}\psi)_{\alpha}(t^b\theta)^i\bigr]\!
\bigl[f^{cde}(\theta^{\dagger} t^d)^j(\bar{\psi}\sigma^{\lambda\sigma})_{\beta}F^c_{\lambda\sigma}\bigr].
\]
\indent
By this means we have shown that the gauge-invariant Lagrangian at quadratic order in the background $\bar{\Psi}^i_{\alpha}$ and $\Psi^i_{\alpha}$ fields possesses rich color and spinor
structure and can contain in principle the terms with an arbitrary power in the strength tensor. In spite of the fact that in this section we restrict ourselves to analysis of the Lagrangian (\ref{eq:5q}), similar reasonings and conclusions can be performed and for the Lagrangian of the form
\[
L_{\bar{\Psi}\bar{\Psi}+\Psi\Psi}=\bar{\Psi}^i_{\alpha}\bar{\Psi}^j_{\beta}
\,\tilde{\cal L}^{ij}_{\alpha\beta}(\theta^{\dagger}\!,\theta,\vartheta)+
\tilde{\cal L}^{\dagger\,ij}_{\alpha\beta}(\theta^{\dagger}\!,\theta,\vartheta)\Psi^i_{\alpha}\Psi^j_{\beta\,}.
\]
\indent
In summary note that we may address a question concerning construction of the gauge-invariant Lagrangian dependent on the background Grassmann-valued spinor fields to an arbitrary power. All these Lagrangians can be thought as separate terms in an expansion of the total Lagrangian if the latter is presented as a formal infinite series
\[
L = \sum\limits_{n=0}^{\infty}\sum\limits_{m=0}^{\infty}
\bar{\Psi}^{i_1}_{\alpha_1}\ldots\bar{\Psi}^{i_n}_{\alpha_n}
{\cal L}^{i_1\,\ldots\, i_n\,j_1\,\ldots\, j_m}_{\alpha_1\ldots\alpha_n\,\beta_1\ldots\beta_m}\!
\Psi^{j_1}_{\beta_1}\ldots\Psi^{j_m}_{\beta_m}\,
+\, {\rm(compl.\,conj.)}.
\]
The coefficient function  ${\cal L}^{i_1\,\ldots\,i_n\,j_1\,\ldots\,j_m}_{\alpha_1\ldots\alpha_n\,\beta_1\ldots\beta_m}$ represents a complicated function of color charges $\theta^{i\dagger},\,
\theta^i$, $\vartheta^a$, gauge field $A^a_{\mu}(x)$ and spinors $\bar{\psi}_{\alpha}$, $\psi_{\alpha}$. The contributions with derivatives of the $\Psi$-fields should be also incorporated in this Lagrangian.

\section{\bf Iteration method for constructing additional currents and sources}
\setcounter{equation}{0}

This section will be devoted to a determination of a regular method for computing additional currents and sources, some of which were considered in the papers \cite{markov_NPA_2007, markov_IJMPA_2010}. We will confine our attention to a system of equations (\ref{eq:2t})\,--\,(\ref{eq:2u}), assuming that the motion of a particle is specified and in the simplest case it represents a straight line
${\bf x} = {\bf v}t$. Furthermore, the $\mbox{parameter}\,\tau$ appearing in equations (\ref{eq:2t})\,--\,(\ref{eq:2u}) in the gauge $e=1/m$ is assumed to be the proper time. For practical
computations, it is more convenient to pass to the coordinate time, as is the case in a system of equations (\ref{eq:1y}), (\ref{eq:1u}). Setting $d\tau = \sqrt{1-{\bf v}^2}\,dt$, instead of (\ref{eq:2t}), now we will have
\[
\frac{d\theta^i(t)}{dt} + igv^{\mu\!}A^a_{\mu}(t,{\bf v}t)(t^a)^{ij}
\theta^j(t)
+ig\bigl(\bar{\chi}_{\alpha}\Psi_{\alpha}^i(t,{\bf v}t)\bigr)
\]
\begin{equation}
-\;ig\varepsilon f_{0\,}Q^a(t)(t^a)^{ij}\bigl(\bar{\chi}_{\alpha}\Psi_{\alpha}^j(t,{\bf v}t)\bigr)
\label{eq:6q}
\end{equation}
\[
-\;ig\varepsilon\hspace{0.025cm}(t^a)^{ij}\theta^{j}(t)\,\Bigl\{\!f_{0\,}\theta^{\dagger{l}}(t)(t^{a})^{lk}\bigl(\bar\chi_{\alpha}\Psi^{k}_{\alpha}(t,{\bf v}t)\bigr)
+f_{0}^{\ast}\bigl(\bar\Psi^{k}_{\alpha}(t,{\bf v}t)\chi_{\alpha}\bigr)(t^{a})^{kl}\theta^{l}(t)\Bigr\}=0
\]
with the initial condition $\left.\theta^i(t)\right|_{\,\!t=t_0}\!=\!\theta^i(t_0)$. Here, we have introduced the spinor $\chi_{\alpha}$ into consideration according to the rule
\[
\chi_{\alpha}\equiv\frac{\sqrt{1-{\bf v}^2}}{\sqrt{2}\,m}\,\psi_{\alpha} =
\frac{1}{\sqrt{2}\,E}\,\psi_{\alpha}.
\]
It is this spinor\footnote{\,It is easy to verify that in this case we correctly reproduce unusual form of density matrix (\ref{eq:1i}) when it is considered that for
fully unpolarized state of a particle, the standard relation is hold \cite{berestetski}
\[
\psi_{\alpha} \bar{\psi}_{\beta} \simeq \frac{1}{2}\,E\,(v\cdot\gamma)_{\alpha\beta}
\]
up to a correction term of $(m/E)$ order (see, footnote in Introduction).}
that appears in equations (\ref{eq:1y}) and (\ref{eq:1u}). Similar equations can be written for color charges $\theta^{\dagger{i}}$ and $Q^a$. The background fields in (\ref{eq:6q})
are given on the straight pass.\\
\indent
Let us rewrite also the expressions for color current (\ref{eq:2p}) and for color source (\ref{eq:2s}) in the representation of coordinate time:
\begin{equation}
j^a_{\mu}(x) = g\hspace{0.03cm}v_{\mu\,}Q^a(t)\,{\delta}^{(3)}({\bf x}-{\bf v}t),
\label{eq:6w}
\end{equation}
\begin{equation}
\eta^i_{\alpha}(x) = g\hspace{0.03cm}\bigl\{\chi_{\alpha\,}\theta^i(t)-
\varepsilon \chi_{\alpha\,}f^{\ast}_{0\,}Q^a(t)(t^a)^{ij\,}\theta^j(t)\bigr\}\,{\delta}^{(3)}({\bf x}-{\bf v}t).
\label{eq:6e}
\end{equation}
In the expression for current (\ref{eq:6w}) we dropped the contribution associated with background fermion field. The parameter $\varepsilon$ we have introduced into (\ref{eq:6q})
and (\ref{eq:6e}) by hands is an effective ``small'' parameter related to terms nonlinear in color charges in the stated expressions (it can be introduced as a factor in the last
term of an initial Lagrangian (\ref{eq:2w})). This parameter is taking as unity at the end of all calculations. Finally, we simplify the problem still further considering that the complex function $f_0$ is independent of time.\\
\indent
We can seek a solution of equation (\ref{eq:6q}) and equations for $\theta^{\dagger{i}}$ and $Q^a$ in the form of an expansion in powers of the parameter $\varepsilon$. It turns out more simple to restrict ourselves to consideration of the solutions of equations only for the $\theta^i(t)$ and $\theta^{\dagger{i}}(t)$ charges, and reproduce the usual charge $Q^a$ with the help of the relation (\ref{eq:2e}). Thus, if we define the solutions in the following form
\begin{equation}
\begin{split}
&\theta^i(t)=\theta^{(0)i}(t)+\varepsilon\hspace{0.03cm}\theta^{(1)i}(t)+\varepsilon^2\theta^{(2)i}(t)+\varepsilon^3\theta^{(3)i}(t)+\,\ldots,\\
&\theta^{\dagger i}(t)=\theta^{\dagger(0){i}}(t)+\varepsilon\hspace{0.03cm}\theta^{\dagger(1){i}}(t)+\varepsilon^2\theta^{\dagger(2){i}}(t)+\varepsilon^3\theta^{\dagger(3){i}}(t)+\,\ldots,\\
\end{split}
\label{eq:6r}
\end{equation}
then
\begin{equation}
Q^a(t)=Q^{(0)a}(t)+\varepsilon\hspace{0.03cm}{Q}^{(1)a}(t)+\varepsilon^2Q^{(2)a}(t)+\varepsilon^3Q^{(3)a}(t)+\,\ldots,\\
\label{eq:6t}
\end{equation}
where
\begin{equation}
Q^{(0)a}(t) =\, \theta^{\dagger(0)}(t)\,t^a\theta^{(0)}(t),
\label{eq:6y}
\end{equation}
\begin{equation}
Q^{(1)a}(t) =\, \theta^{\dagger(0)}(t)\,t^a\theta^{(1)}(t) +\,\theta^{\dagger(1)}(t)\,t^a\theta^{(0)}(t),
\label{eq:6u}
\end{equation}
\[
Q^{(2)a}(t) =\, \theta^{\dagger(1)}(t)\,t^a\theta^{(1)}(t) + \bigl\{\theta^{\dagger(0)}(t)\,t^a\theta^{(2)}(t) + \,\theta^{\dagger(2)}(t)\,t^a\theta^{(0)}(t)\bigr\}
\]
and so on. It is worth noting that in (\ref{eq:6r}) and (\ref{eq:6t}) each term of the expansion is gauge-covariant {\it irrespective} the others. As a result, under the substitution of (\ref{eq:6r}), (\ref{eq:6t}) in (\ref{eq:6w}) or (\ref{eq:6e}) to each order in $\varepsilon$ we will have a new gauge-covariant current or source.\\
\indent
Let us substitute expansion (\ref{eq:6r}) into equation (\ref{eq:6q}). Correct to first order in $\varepsilon$, we obtain
\[
\begin{split}
&\frac{d\theta^{(0)i}(t)}{dt} + ig\hspace{0.03cm}v^{\mu\!}A^a_{\mu}(t,{\bf v}t)(t^a)^{ij}\theta^{(0)j}(t)
+ig\bigl(\bar{\chi}_{\alpha}\Psi_{\alpha}^i(t,{\bf v}t)\bigr)=0,\\
&\frac{d\theta^{(1)i}(t)}{dt} + ig\hspace{0.03cm}v^{\mu\!}A^a_{\mu}(t,{\bf v}t)(t^a)^{ij}\theta^{(1)j}(t)
-igf_{0\,}Q^{(0)a}(t)(t^a)^{ij}\bigl(\bar{\chi}_{\alpha}\Psi_{\alpha}^j(t,{\bf v}t)\bigr)\\
-\,&ig\hspace{0.025cm}(t^a)^{ij}\theta^{(0)j}(t)\Bigl\{\!f_{0\,}\theta^{\dagger(0){l}}(t)(t^{a})^{lk}\bigl(\bar\chi_{\alpha}\Psi^{k}_{\alpha}(t,{\bf v}t)\bigr)
+f_{0}^{\ast}\bigl(\bar\Psi^{k}_{\alpha}(t,{\bf v}t)\chi_{\alpha}\bigr)(t^{a})^{kl}\theta^{(0)l}(t)\Bigr\}=0.
\end{split}
\]
Solution of the first equation has the following structure:
\begin{equation}
\theta^{(0)i}(t)=\theta^{i}_{0}(t) + \Omega^i(t),\quad \theta^{i}_{0}(t)\equiv U^{ij}(t,t_0)\theta^{j}(t_0),
\label{eq:6i}
\end{equation}
where the evolution operator in the fundamental representation $U(t, t_0)$ is defined by expression (\ref{eq:5y}), and an explicit form of function $\Omega^{i}(t)$
(and its conjugation $\Omega^{\dagger{i}}(t)$) is given in $\mbox{Appendix\,A,}$ Eq.\,(A.1). Inserting (\ref{eq:6i}) into (\ref{eq:6y}), we derive further
\begin{equation}
Q^{(0)a}(t) = Q^{a}_0(t) + \bigl\{\theta^{\dagger}_0(t)\,t^a\Omega(t) + \,\Omega^{\dagger}(t)\,t^a\theta_0(t)\bigr\} +\,
\Omega^{\dagger}(t)\,t^a\Omega(t),
\label{eq:6o}
\end{equation}
where
\[
Q^{a}_0(t)\equiv\tilde{U}^{ab}(t,t_0)Q^{b}(t_0)
\]
and in turn
\[
\tilde{U}(t,t_0)={\rm T}\exp\Biggl\{-ig\!\int\limits_{t_0}^t
\!\bigl(v\cdot A^a(t^{\prime},{\bf v}t^{\prime})\bigr)T^a dt^{\prime}\!\Biggr\}
\]
is the evolution operator in the adjoint representation. In deriving the expression $Q^a_0(t)$ we have taken into account the identity
\begin{equation}
U(t_0,t)\,t^{a\,}U(t, t_0)=\tilde{U}^{ab}(t,t_0)\,t^b.
\label{eq:6p}
\end{equation}
If we substitute (\ref{eq:6o}) and (\ref{eq:6i}) into (\ref{eq:6w}), (\ref{eq:6e}), then to the zeroth-order approximation in $\varepsilon$, we properly reproduce
the simplest current and source written out in Section 5 of \cite{markov_NPA_2007} (Eq.\,(A.2) in Appendix A of the present work).\\
\indent
Furthermore, a solution of the equation for $\theta^{(1)i}(t)$ has the following structure:
\begin{equation}
\theta^{(1)i}(t)=igf_{0\!}\!\int\limits_{t_0}^t\!U^{ij}(t,t^{\prime})Q^{(0)a}(t^{\prime})(t^a)^{jk}
\bigl(\bar{\chi}_{\alpha}\Psi_{\alpha}^k(t^{\prime},{\bf v}t^{\prime})\bigr)\,dt^{\prime}
\label{eq:6a}
\end{equation}
\[
\hspace{0.7cm}
\begin{split}
+\;ig\!\!\int\limits_{t_0}^t\!U^{ij}(t,t^{\prime})(t^a)^{js}\theta^{(0)s}(t^{\prime})\,
\Bigl\{&f_{0\,}\theta^{\dagger(0){l}}(t^{\prime})(t^{a})^{lk}
\bigl(\bar\chi_{\alpha}\Psi^{k}_{\alpha}(t^{\prime},{\bf v}t^{\prime})\bigr)\\
&+f_{0}^{\ast}\bigl(\bar\Psi^{k}_{\alpha}(t^{\prime},{\bf v}t^{\prime})\chi_{\alpha}\bigr)(t^{a})^{kl}
\theta^{(0)l}(t^{\prime})\Bigr\}\,dt^{\prime}.
\end{split}
\]
It will be shown below that ``correction'' (\ref{eq:6a}) (and a similar expression for $\theta^{\dagger(1)i}$) encloses all additional sources (A.4)\,--\,(A.7), and taking into account
relation (\ref{eq:6u}), properly reproduces the additional current (A.3). For this purpose we substitute the solutions (\ref{eq:6i}) and (\ref{eq:6o}) into (\ref{eq:6a}) and
retain only terms linear in the charges $\theta_0^i(t)$ and $Q_0^a(t)$. After some tedious manipulations employing the identity (\ref{eq:6p}), expression
(\ref{eq:6a}) can be cast into the following form:
\[
\theta^{(1)i}(t)|_{\rm linear\,in\,\theta_0,\,{\it Q}_0}= - {\rm Re}(f_0)\Bigl\{
Q^a_0(t)(t^a\Omega(t))^i +
\bigl[\,\Omega^{\dagger}(t)\hspace{0.035cm}t^a\Omega(t)\bigr](t^a\theta_0(t))^i
\]
\[
+\,\bigl[\,\theta_0^{\dagger}(t)\hspace{0.035cm}t^a\Omega(t) + \Omega^{\dagger}(t)\hspace{0.035cm}t^a\theta_0(t)\bigr](t^a\Omega(t))^i
\Bigl\}
\]
\begin{equation}
-\,i\,{\rm Im}(f_0)\biggl\{
Q^a_0(t)(t^a\Omega(t))^i
+\bigl[\,\theta_0^{\dagger}(t)\hspace{0.035cm}t^a\Omega(t)\bigr](t^a\Omega(t))^i
\hspace{1cm}
\label{eq:6s}
\end{equation}
\[
\begin{split}
+\! &\int\limits_{t_0}^t\!\bigl[\hspace{0.035cm}\theta_0^{\dagger}(t)\hspace{0.035cm}t^a\Phi(t,t^{\prime}) + \Phi^{\dagger}(t,t^{\prime})\,t^a\theta_0(t)\bigr]\frac{(t^a\Phi(t,t^{\prime}))^i}{dt^{\prime}}\,dt^{\prime}\\
+\! &\int\limits_{t_0}^t\biggl[\theta_0^{\dagger}(t)\,t^a\frac{\Phi(t,t^{\prime})}{dt^{\prime}} -
\frac{\Phi^{\dagger}(t,t^{\prime})}{dt^{\prime}}\,t^a\theta_0(t)\biggr](t^a\Phi(t,t^{\prime}))^i\hspace{0.035cm}dt^{\prime}\\
+\! &\int\limits_{t_0}^t\biggl[\Phi^{\dagger}(t,t^{\prime})\,t^a\frac{\Phi(t,t^{\prime})}{dt^{\prime}} -
\frac{\Phi^{\dagger}(t,t^{\prime})}{dt^{\prime}}\,t^a\Phi(t,t^{\prime})\biggr]dt^{\prime}\,(t^a\theta(t))^i
\biggr\}.
\end{split}
\]
Here, we have considered the function
\[
\Phi(t,t^{\prime})\equiv U(t,t^{\prime})\,\Omega(t^{\prime})
\]
and used its following properties obvious from the definition:
\[
\frac{\Phi(t,t^{\prime})}{dt^{\prime}}=-ig\,U(t,t^{\prime})
\bigl(\bar{\chi}_{\alpha}\Psi_{\alpha}(t^{\prime},{\bf v}t^{\prime})\bigr),\quad
\Phi(t,t)=\Omega(t),\quad
\Phi(t,t_0)=0.
\]
A surprising feature of the expression obtained (\ref{eq:6s}) is that all the integrands in the terms proportional to ${\rm Re}f_0$ grouped together in the total differential. This has allowed us
to perform easily the integration in $t^{\prime}$.\\
\indent
Furthermore, we will consider the first `correction' to the initial source
$$
\eta^{(0)i}_{\alpha}(x)=g\hspace{0.025cm}\chi_{\alpha}\hspace{0.035cm}\theta^{(0)i}(t)\,{\delta}^{(3)}({\bf x}-{\bf v}t),
$$
which by virtue of (\ref{eq:6e}) has the following form:
\[
\eta^{(1)i}_{\alpha}(x) = g\bigl\{\chi_{\alpha\,}\theta^{(1)i}(t)-
\chi_{\alpha\,}f^{\ast}_{0\,}Q^{(0)a}(t)(t^a)^{ij\,}\theta^{(0)j}(t)\bigr\}\,{\delta}^{(3)}({\bf x}-{\bf v}t).
\]
Let us substitute expressions (\ref{eq:6s}), (\ref{eq:6o}) and (\ref{eq:6i}) for $\theta^{(1)i}(t)$, $Q^{(0)a}(t)$ and $\theta^{(0)j}(t)$,
respectively, and retain only the terms linear in $Q^a_0(t)$ and $\theta^j_0(t)$. Then we have
\begin{equation}
\eta^{(1)i}_{\alpha}(x) = -2g\,{\rm Re}(f_{0})\chi_{\alpha}Q^a_0(t)(t^a)^{ij\,}\Omega^j(t)\,{\delta}^{(3)}({\bf x}-{\bf v}t)
\label{eq:6d}
\end{equation}
\[
\begin{split}
&-2g\,{\rm Re}(f_{0})\chi_{\alpha\!}\bigl[\,\theta^{\dagger}_0(t)\,t^a\Omega(t) + \,\Omega^{\dagger}(t)\,t^a\theta_0(t)\bigr](t^a)^{ij\,}\Omega^j(t)\,{\delta}^{(3)}({\bf x}-{\bf v}t)\\
&-2g\,{\rm Re}(f_{0})\chi_{\alpha\!}\bigl[\,\Omega^{\dagger}(t)\,t^a\Omega(t)\bigr](t^a)^{ij\,}\theta^j_0(t)\,{\delta}^{(3)}({\bf x}-{\bf v}t)\\
\end{split}
\]
\[
-\,(\mbox{the terms proportional to ${\rm Im}f_0$}).
\hspace{1.3cm}
\]
This source should be compared with additional ones (A.4)\,--\,(A.7). Comparing the first, second and third terms in (\ref{eq:6d}) with expressions (A.4)\,--\,(A.7), we obtain
\begin{equation}
\alpha=\beta=\beta_1=\tilde{\beta}_1=-\hspace{0.015cm}2\hspace{0.02cm}{\rm Re}f_0.
\label{eq:6f}
\end{equation}
\indent
Now we turn to calculating the correction $Q^{(1)a}(t)$. At first we rewrite expression (\ref{eq:6u}) substituting explicitly the solution $\theta^{(0)i}(t)$ (\ref{eq:6i}).
Here, we have
\begin{equation}
Q^{(1)a}(t) = \bigl[\, \theta^{\dagger}_0(t)\,t^a\theta^{(1)}(t) +\,\theta^{\dagger(1)}(t)\,t^a\theta_0(t)\bigr]
+
\bigl[\,\Omega^{\dagger}(t)\,t^a\theta^{(1)}(t) +\,\theta^{\dagger(1)}(t)\,t^a\Omega(t)\bigr].
\label{eq:6g}
\end{equation}
In the usual manner, we restrict our consideration only to contributions linear in $\theta_0^i(t)$ and $Q_0^a(t)$. By virtue of this fact, one can substitute the above-obtained expression (\ref{eq:6s}) into the last two terms in (\ref{eq:6g}) for $\theta^{(1)i}(t)$. However, for the first and second terms in (\ref{eq:6g}) this substitu\-tion leads to the second order terms in the color charges $\theta_0^i(t)$ and $Q_0^a(t)$. Therefore, here we need to return to the initial expression (\ref{eq:6a}) and to single out the contributions completely independent of the color charges $\theta_0^i(t)$ and $Q_0^a(t)$. For this purpose, it is merely necessary to substitute the last terms of the solutions (\ref{eq:6i}) and (\ref{eq:6o}) into expression (\ref{eq:6a}) for $\theta^{(0)i}(t^{\prime})$ and $Q^{(0)a}(t^{\prime})$. Straightforward calculations result in
\begin{equation}
\theta^{(1)i}(t)|_{\rm free\,in\,\theta_0,\,{\it Q}_0}= -\hspace{0.015cm}{\rm Re}(f_0)\bigl[\,\Omega^{\dagger}(t)\hspace{0.035cm}t^{a\hspace{0.025cm}}\Omega(t)\bigr](t^{a}\Omega(t))^i
\label{eq:6h}
\end{equation}
\[
\begin{split}
&-i\,{\rm Im}(f_0)\Biggl\{\,
\int\limits_{t_0}^t\!\bigl[\,\Phi^{\dagger}(t,t^{\prime})\hspace{0.035cm}t^a\Phi(t,t^{\prime})\bigr]\frac{(t^a\Phi(t,t^{\prime}))^i}{dt^{\prime}}\,dt^{\prime}\\
+\! &\int\limits_{t_0}^t\biggl[\,\Phi^{\dagger}(t,t^{\prime})\,t^a\frac{\Phi(t,t^{\prime})}{dt^{\prime}} -
\frac{\Phi^{\dagger}(t,t^{\prime})}{dt^{\prime}}\,t^a\Phi(t,t^{\prime})\biggr](t^a\Phi(t,t^{\prime}))^i dt^{\prime}
\Biggr\}.
\end{split}
\]
Here we observe the same picture as in (\ref{eq:6s}). The integrands in the terms proportional to ${\rm Re}f_0$ grouped together in the total differential. The consequence of this fact is such a simple form of the first term in (\ref{eq:6h}).\\
\indent
Substituting (\ref{eq:6h}) into the first and second terms of expression (\ref{eq:6g}) for $\theta^{(1)}(t)$, and (\ref{eq:6s}) into the third and forth terms of the same expression, we find from (\ref{eq:6w}) the first correction to the initial current $j_{\mu}^{(0)a}(x)$:
\begin{equation}
j^{(1)a}_{\mu}(x)=-\hspace{0.015cm}g\hspace{0.025cm}{\rm Re}(f_0)\hspace{0.03cm}v_{\mu\,}Q^b_0(t)\bigl[\,\Omega^{\dagger}(t)\,\{t^a\!,t^{b}\}\,\Omega(t)\bigr]{\delta}^{(3)}({\bf x}-{\bf v}t)
\label{eq:6j}
\end{equation}
\[
\begin{split}
&-g\hspace{0.025cm}{\rm Re}(f_0)\hspace{0.03cm}v_{\mu}
\bigl[\,\Omega^{\dagger}(t)\,t^{b\,}\theta_0(t) + \theta^{\dagger}_{0\,}t^{b\,}\Omega(t) \bigr]
\!\bigl[\,\Omega^{\dagger}(t)\,\{t^a\!,t^{b}\}\,\Omega(t)\bigr]{\delta}^{(3)}({\bf x}-{\bf v}t)\\
&-g\hspace{0.025cm}{\rm Re}(f_0)\hspace{0.03cm}v_{\mu}
\bigl[\,\Omega^{\dagger}(t)\,t^{b\,}\Omega(t)\bigr]
\!\bigl[\,\Omega^{\dagger}(t)\,\{t^a\!,t^{b}\}\,\theta_0(t) + \theta^{\dagger}_0\,\{t^a\!,t^{b}\}\,\Omega(t) \bigr]
{\delta}^{(3)}({\bf x}-{\bf v}t)
\end{split}
\]
\[
-\,(\mbox{the terms proportional to ${\rm Im}f_0$}).
\]
Here, the first term on the right-hand side reproduces the additional current (A.3), if we set
\[
\sigma = {\rm Re}f_0.
\]
\indent
Comparing the last expression with (\ref{eq:6f}), we find that the constants $\sigma$ and $\alpha$ are connected among themselves by relation (A.8), as was obtained in \cite{markov_NPA_2007} from a fundamentally different approach. The second and third terms on the right-hand side of (\ref{eq:6j}) represent new additional currents induced by a moving color particle, which has been overlooked in the above-mentioned paper. In this way from all aforesaid it follows that if we believe the constant $f_0$ to be a {\it pure real}
and set by virtue of (\ref{eq:6f}) and (A.9)
\[
{\rm Re}f_0 = \frac{1}{2}\,\frac{C_F}{T_F}\,,
\]
then the obtained expressions for the first correction to the initial color source and current, Eqs.\,(\ref{eq:6d}) and (\ref{eq:6j}), exactly reproduce additional sources and currents obtained earlier on the basis of heuristic reasoning. This provides a rather strong argument for the correctness of a choice of the initial model Lagrangian (\ref{eq:2w}).

\section{\bf Conclusion}
\setcounter{equation}{0}

\begin{figure}[hbtp]
\begin{center}
\begin{tabular}{ccccc}
\includegraphics[width=0.31\textwidth]{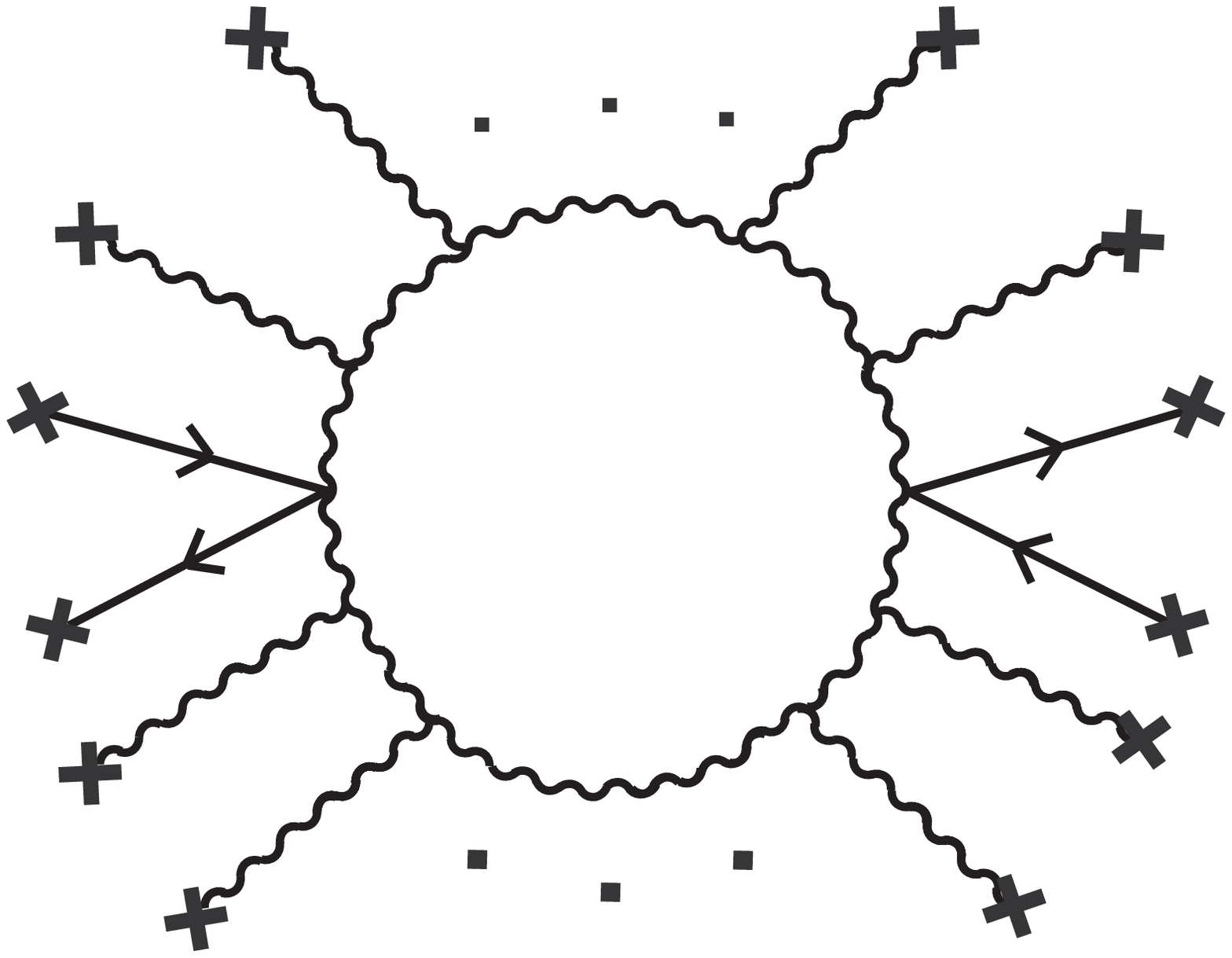}&\hspace{0.05\textwidth}&
\includegraphics[width=0.31\textwidth]{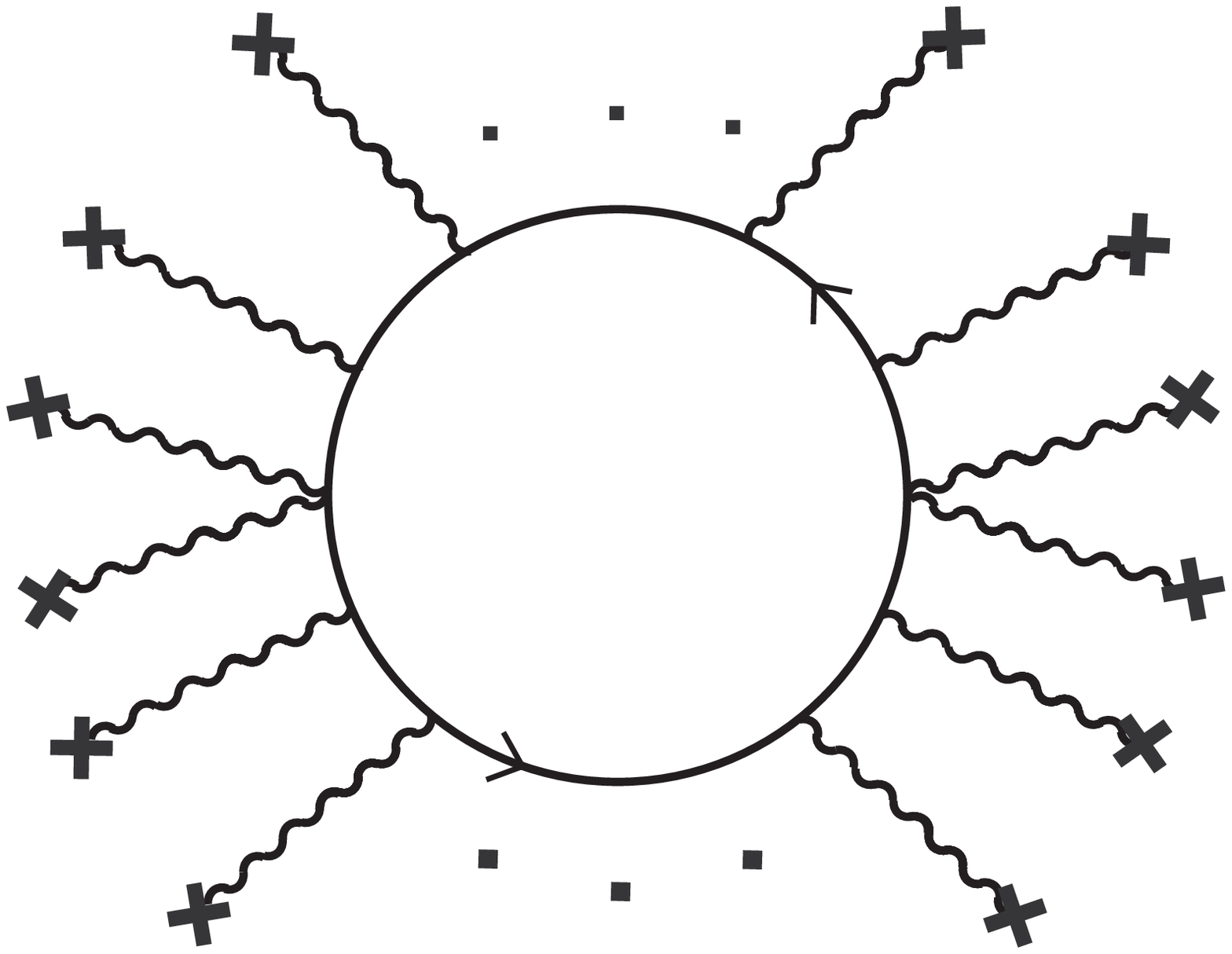}&\hspace{0.05\textwidth}&
\includegraphics[width=0.31\textwidth]{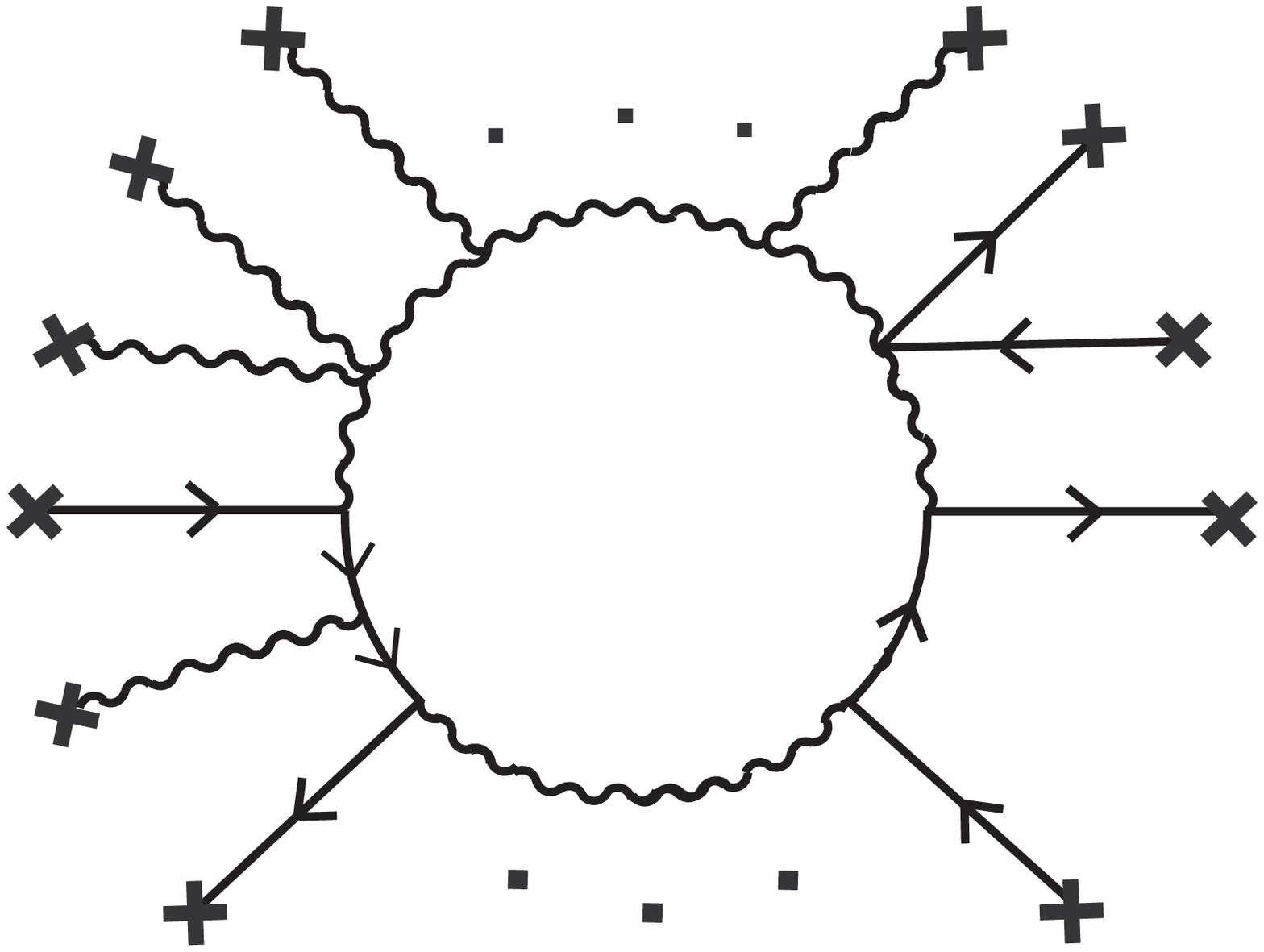}\\
        (a) && (b) && (c)
\end{tabular}
\end{center}
\caption{\small Scattering of a hard particle running in the loop by the background fermionic and bosonic fields.}
\label{fig1}
\end{figure}

Being based only on the general principles, specified in the Introduction, we have considered a problem of the construction of the action that
would describe the dynamics of (pseudo)classical color particles both in background non-Abelian bosonic and fermionic fields.
The major requirement in this construction is the gauge invariance requirement of the action in question under the gauge transformation of
all dynamical variables including the background fields. But here, however, another point arises: the gauge invariance is necessary of course, but
is it sufficient in this case? As has been shown explicitly in Sections 4 and 5, the answer to this question is generally
negative. It has appeared that in principle the existence of an infinite number of contributions to the interaction Lagrangian, which leave the
action real, gauge and reparametrization invariant, is possible. Here, based only on these general principles, there is no way to say
which of these contributions to the interaction Lagrangian actually concern the real dynamics of a color particle, and which are not. Moreover, also there is no way of
fixing the arbitrariness in the scalar `weighting' functions (see Section 4) with which these contributions enter to the Lagrangian.
The last circumstance, in particular, leads to the fact that it is necessary to invoke an additional information for determining these functions. For
a comprehensive Lagrangian description of the dynamics of the color particle in background non-Abelian Bose- and Fermi-fields and rigorous justification of the
results obtained, we come up against the problem of derivation of the Lagrangian from the first principles within the framework of quantum field theory. Obtaining
the Wong equation for the usual color charge in \cite{jalilian_2000} and the equations of motion for the Grassmann color charges in background gauge field in
\cite{borisov_1982, d_hoker_1996, fresneda_2008}, provides examples of such a derivation directly from underlying quantum field theory. These equations can be justified
as a semiclassical approximation to the worldline formula\-tion of the one-loop effective action in QCD. However, an attempt at a direct inclusion of background fermion
field into the developed approaches encounters severe difficulties both technical and fundamental nature.\\
\indent
It was shown in a considerable amount of papers that the one-loop effective actions for scalar models, QED and QCD, could be expressed in terms of a quantum mechanical
path integral over a point particle Lagrangian. For the case of the QCD coupling, the worldline path integral representation was obtained not only for the effective action
for quark loop, but for gluon one in an external non-Abelian field as well \cite{strassler_1992, reuter_1997}. One of the important steps here was made by Borisov and Kulish \cite{borisov_1982},
D'Hoker and Gagn\'e \cite{d_hoker_1996}. They have presented the internal color degrees of freedom in terms of worldline fermions expressed by independent dynamical Grassmann variables $\theta^{\dagger{i}}(t)$ and $\theta^{i}(t)$. It is precisely these color charges that we have used throughout the present paper. They have been first introduced in \cite{barducci_1977, balachandran_1977} from completely different reasoning by means of less-formal considerations. For a rigorous derivation of the dynamical equations presented in Sections 2 and 3, and also of all possible additional contributions to these equations from the terms written out in Sections 4 and 5, it is necessary to consider a more general problem: the worldline path integral representation of the effective one-loop QCD action in the presence of both classical external bosonic and fermionic fields. In terms of a functional superdeterminant the given effective action has the form
\begin{equation}
i\Gamma[A,\bar{\Psi},\Psi]=-\frac{1}{2}\,\ln {\rm SDet}
\left(
\begin{array}{cc}
{\cal D}^{ab}_{\mu\nu}(A,\bar{\Psi},\Psi) & \bar{\cal F}^{\,aj}_{\mu\beta}(A,\bar{\Psi}) \\
{\cal F}^{\,ib}_{\alpha\nu}(A,\Psi) & {\cal D}^{\,ij}_{\alpha\beta}(A)
\end{array}
\right).
\label{eq:7q}
\end{equation}
An explicit form of operators appearing in the supermatrix on the right-hand side of (\ref{eq:7q}) is written out in Appendix B, Eqs.\,(B.3)\,--\,(B.5).
Here, we only note that in deriving the above expression, we have used the second-order formalism for fermions \cite{morgan_1995} instead of the standard Dirac formalism.
To a limited extent, the complexity of expression (\ref{eq:7q}) may be imagined if to rewrite the superdeterminant in terms of ordinary determinants as \cite{berezin_book_1983}
\begin{equation}
i\Gamma[A,\bar{\Psi},\Psi]=
-\frac{1}{2}\ln {\rm Det}\,{\cal D}^{ab}_{\mu\nu}(A,\bar{\Psi},\Psi) +
\frac{1}{2}\ln {\rm Det}\,{\cal D}^{\,ij}_{\alpha\beta}(A)
\label{eq:7w}
\end{equation}
\[
-\;\frac{1}{2}\ln {\rm Det}\bigl(\,{\rm I} - {\cal D}^{-1\,ac}_{\mu\lambda}(A,\bar{\Psi},\Psi)\,\bar{\cal F}^{\,ci}_{\lambda\alpha}(A,\bar{\Psi})\,
{\cal D}^{-1\,ij}_{\alpha\beta}(A)\,{\cal F}^{\,jb}_{\beta\nu}(A,\Psi)\bigr).
\]
The first term on the right-hand side represents the usual one-loop gauge boson effective action. The one-loop determinant may be thought of as being built up by summing over multiple insertions
of the backgrounds, Fig.\,\ref{fig1}(a). Here, unusual insertions of external fermion lines are connected with the additional contributions to the gluon kinetic operator
${\cal D}_{\mu\nu}^{ab}(A,\bar{\Psi},\Psi)$, Eq.\,(B.3), proportional to the background $\Psi$-fields. The appearance of such insertions is a direct consequence of using the second order formalism for fermions. The second term in (\ref{eq:7w}) is the contribution to the gauge boson effective action from dynamical fermions, Fig.\,\ref{fig1}(b). Finally, the last term in (\ref{eq:7w}) gives the effective action for fermions properly interesting for us. Formally, here under the sign of determinant there is an expression, which is nonlocal and substantially nonlinear with respect to the background gauge field. Besides, since the $\Psi$-fields are explicitly involved into the gluon kinetic operator, then the expression under the sign of determinant formally also contains the background fermion fields to an arbitrary (even) power. The last circumstance distinguishes considerably the given effective action from a similar action in the standard first-order formalism \cite{elmfors_1999}, where under the sign of determinant we have only one pair the $\Psi$-fields.\\
\indent
It appears very difficult, if possible at all in this situation, to define the worldline path integral representation for the effective action\footnote{\,In fact it is
this circumstance that serves a basic motivation for writing the present work and the next one \cite{part_II}. Within this crude approach we would like to make clear which of the contributions
can appear in the required action and what is their structure. Such a preliminary work, as we hope, enables us at least on qualitative level to see what we should expect in a more rigorous approach and
in some extent to facilitate the construction of a comprehensive theory.}. Perhaps, the required representation should be constructed at once beginning with superdeterminant (\ref{eq:7q}), without reducing it to usual determinants. However, it is not clear so far how this can be made within the developed techniques \cite{borisov_1982, d_hoker_1996, fresneda_2008, strassler_1992, reuter_1997}. To the best of our knowledge, the given problem has not been considered in the literature. In fact, the various contributions to the interaction Lagrangian derived in the present work can be considered as `fragments' of the total Lagrangian that would enter into the worldline path integral representation of the effective action $\Gamma[A,\bar{\Psi},\Psi]$.\\
\indent
It may be taken a different view of the given problem. The presence of the fermionic background field leads to qualitatively new phenomenon: a single background fermion can change a
particle\footnote{\,The first-quantized field theory considers a particle in a loop as a single entity.} in the loop from a Dirac spinor into a vector boson and vice versa, Fig.\,\ref{fig1}(c).
Therefore, our purpose is to construct a theory which consistently describes a particle that can be either vector boson or Dirac fermion. From the mathematical point of view this means that it is necessary to find  an explicit form of the fermion vertex operator which is inserted into the world closed line of the hard particle and defines the radiation (or absorption) process of an external quark simulated by an external fermionic background. The construction of this vertex operator is a necessary ingredient for a rigorous derivation of the evolution equations for color charges.
One way of looking for a solution of the problem at hand (and in particular of computa\-tion of the desired vertex operator) is through the string theory. At one time in a number of works
\cite{lovelace_1984, fradkin_1985, sen_1985} the problem of the propagation of (super)string in background fields, was considered. As shown in \cite{callan_1985}, background spacetime
fermions may be incorporated into the string action on equal terms with the other external fields if to use the covariant string vertex operator \cite{knizhnik_1985, friedan_1986}.
As far as we know, this is the only rigorous inclusion of interaction with an external fermion field which is well understood. Here, the following heuristic argument is applicable: the required
fermion vertex for the first-quantized field theory is related, in a certain way, to the fermion vertex operator\footnote{\,One of the indirect proofs of the existence of such a relation is the fact that there exists practically perfect coincidence in a structure between boson vertex operator in string theory and boson vertex operator arising in considering the effective actions for spinor and vector boson particles in background gauge field \cite{strassler_1992}.} of superstring theory. The efficiency of the string-based methods in concrete applications to the problems of calculation of the pure gluon one-loop QCD amplitudes was demonstrated in the early 1990s by Bern and Kosower \cite{bern_1992} and then by the others. Now our purpose is to extend the well-developed
approach\footnote{\,Note that the authors of \cite{bern_1992} planned to consider this more general case, but here, they used the usual field-theoretical approach \cite{bern_1995}.} to incorporate external quarks.\\
\indent
Thus, in light of this, one can outline another way of derivation of the color charge evolution equations from the first principles. Our first task here is to define in an explicit form the effective one-loop QCD amplitude including both external bosons and external fermions in the context of superstring-inspired approach \cite{bern_1992}. Then the second task is to guess an explicit form of the effective action in the worldline formulation which, on expanding in powers of the background fields, would reproduce the mixed quark-gluon one-loop amplitudes obtained
at the first stage. And the final step would be the worldline representation for the color degree of freedom of hard particle running in the mixed loop in the spirit of Borisov and Kulish \cite{borisov_1982} or D'Hoker and Gagn\'e \cite{d_hoker_1996}.\\
\indent
Finally, one can look at the solution of this problem purely from a geometrical point of view. In the paper by Duval and Horvathy \cite{duval_1982} within (pre)simplectic geometry an alternative deriving Wong's equation have been given. Furthermore, Horvathy \cite{horvathy_1982} have shown that the developed approach, which is in fact a sort of generalized variational calculus, is basically equivalent to that of Barducci {\it et al} \cite{barducci_1977} and Balachandran {\it et al} \cite{balachandran_1977}. It would be very interesting to extend this geometrical approach to the case of the presence of Grassmann-valued background fields.

%%%%%%%%%%%%%%%%%%%%%%%%%%%%%%%%%%%%%%%%%%%%%%%%%%%%%%%%%%%%%%%%%%%%%%%%%%%%%%%%%%%%%%%%%%%%%%%%%%%%%%%%%%%%%%%%%%%%%%%%%%%%%%%%%%%%%%%%%%%%%%%%%%%%%%%%%%%%%%%%%%%%%%%%%%%%%%%%%%%%%%%%%%%%%%%

\section*{\bf Acknowledgments}
The authors are grateful to Peter Horvathy for correspondence.
This work was supported by the Russian Foundation for Basic Research (project No. 09-02-00749),
by the grant of the President of Russian Federation for the support of the Leading
Scientific Schools (NSh-1027.2008.2), in part by the Federal Target Programs "Development of Scientific Potential in
Higher Schools" (project 2.2.1.1/1483, 2.1.1/1539), and "Research and Training Specialists in Innovative Russia, 2009-2013",
contract 02.740.11.5154.

\newpage
\section*{\bf Appendix A}
\setcounter{equation}{0}

In this Appendix we give an explicit form of additional gauge-covariant currents and sources suggested in \cite{markov_NPA_2007, markov_IJMPA_2010}. For convenience
of notations of these currents and sources, we introduce the following functions:
$$
\Omega^i(t)\equiv
-ig\!\!\int\limits_{t_0}^{t}\!
U^{ij}(t,t^{\prime})
\bigl(\bar{\chi}_{\alpha}\Psi^{j}_{\alpha}(t^{\prime},{\bf v}t^{\prime})\bigr)\,dt^{\prime},
\quad
\Omega^{\dagger i}(t) \equiv
ig\!\!\int\limits_{t_0}^{t}
\bigl(\bar{\Psi}^{j}_{\alpha}
(t^{\prime},{\bf v}t^{\prime})\chi_{\alpha}\bigr)
U^{ji}(t^{\prime},t)\,dt^{\prime}.
\eqno{\rm (A.1)}
$$
In the terms of these functions two additional currents in \cite{markov_NPA_2007} have the following form:
$$
j_{\theta\mu}^{\,a}(x) = g\hspace{0.015cm}v_{\mu}\bigl\{\Omega^{\dagger j}(t)(t^a)^{j\hspace{0.01cm}i}\theta^i_0(t)+
\theta^{\dagger{i}}_0(t)(t^a)^{ij\,}\Omega^j(t)\bigr\}\,{\delta}^{(3)}({\bf x}-{\bf v}t),
\eqno{\rm (A.2)}
$$
$$
j_{\Xi\,\mu}^{\,a}(x) = \sigma g\hspace{0.015cm}v_{\mu}Q^b_0(t)\bigl[\Omega^{\dagger i}(t)\{t^a\!,t^b\}^{ij\,}\Omega^j(t)\bigr]
{\delta}^{(3)}({\bf x}-{\bf v}t),
\hspace{1.6cm}
\eqno{\rm (A.3)}
$$
where $\theta^i_0(t)=U^{ij}(t,t_0)\,\theta^j(t_0)$ and $Q^a_0(t)=\tilde{U}^{ab}(t,t_0)\,Q^b_0(t_0)$.\\
\indent
Furthermore, we give a list of additional sources
$$
\eta_{Q\alpha}^i(x) = \alpha\hspace{0.025cm}g\chi_{\alpha\,}Q^a_0(t)(t^a)^{ij\,}\Omega^j(t)\,{\delta}^{(3)}({\bf x}-{\bf v}t),
\eqno{\rm (A.4)}
$$
$$
\eta_{\Omega\,\alpha}^i(x)=\beta_{1}g\,\chi_{\alpha}
(t^a)^{ij}\,\Omega^{j}(t)
\bigl[\,\Omega^{\dagger{k}}(t)(t^a)^{kl}\hspace{0.03cm}\theta^l_0(t)\bigr]{\delta}^{(3)}({\bf x}-{\bf v}t),
\eqno{\rm (A.5)}
$$
$$
\eta_{\tilde{\Omega}\,\alpha}^i(x)=\tilde{\beta}_{1}g\,\chi_{\alpha}
(t^a)^{ij}\,\Omega^{j}(t)
\bigl[\,\theta^{\dagger{l}}_0(t)(t^a)^{lk}\hspace{0.03cm}\Omega^{k}(t)\bigr]{\delta}^{(3)}({\bf x}-{\bf v}t),
\hspace{0.2cm}
\eqno{\rm (A.6)}
$$
$$
\eta_{\Xi\,\alpha}^i(x)=\beta g\,\chi_{\alpha}(t^a)^{ij}\,\theta^j_0(t)
\bigl[\,\Omega^{\dagger{k}}(t)(t^a)^{kl}\hspace{0.03cm}\Omega^{l}(t)\bigr]
{\delta}^{(3)}({\bf x}-{\bf v}t).
\hspace{0.3cm}
\eqno{\rm (A.7)}
$$
In \cite{markov_NPA_2007} we have also obtained a relation between constants in expressions (A.3) and (A.4), and also an explicit value of one of them
in terms of the group invariants:
$$
{\rm Re}\,\sigma=\frac{1}{2}\,\alpha\,,
\eqno{\rm (A.8)}
$$
$$
\alpha =-\frac{C_F}{T_F}\,.
\eqno{\rm (A.9)}
$$

%%%%%%%%%%%%%%%%%%%%%%%%%%%%%%%%%%%%%%%%%%%%%%%%%%%%%%%%%%%%%%%%%%%%%%%%%%%%%%%%%%%%%%%%%%%%%%%%%%%%%%%%%%%%%%%%%%%%%%%%%%%%%%%%%%%%%%%%%%%%%%%%%%%%%%%%%%%%%%%%%%%%%%%%%%%%%%%%%%%%%%%%%%%%%%
\newpage

\section*{\bf Appendix B}
\setcounter{equation}{0}

Let us write down an initial expression of the classical QCD action
$$
S[\,A,\bar{\Psi},\Psi] = -\frac{1}{4}\int\!d^{\,4}x\hspace{0.025cm}F^a_{\mu\nu}(x)F^{a\hspace{0.013cm}\mu\nu}(x)
\eqno{\rm (B.1)}
$$
$$
+\!\int\!d^4x\,\bar{\Psi}(x)\Bigl\{D_{\mu}(A)D^{\mu}(A) - \frac{1}{2}\,gF_{\mu\nu}(x)\sigma^{\mu\nu} + m^2\Bigr\}\Psi(x) + \,\ldots\,.
$$
Here, $D_{\mu}(A)=\partial_{\mu}+igA^a_{\mu}(x)t^a$, $\sigma^{\mu\nu}=[\gamma^{\mu},\gamma^{\nu}]/2i$, $F_{\mu\nu}(x)=F_{\mu\nu}^a(x)t^a$ and the ellipsis is refereed to the gauge fixing and ghost terms. The action for fermions is written down in the representation of the second-order formalism. To provide the effective action, according to the standard procedure \cite{weinberg_book_2001} in the first step we replace the fields in (B.1) by the shifted ones:
$$
A^a_{\mu}(x) \rightarrow A^a_{\mu}(x) + a^a_{\mu}(x),\;
\bar{\Psi}^i_{\alpha}(x) \rightarrow \bar{\Psi}^i_{\alpha}(x) + \bar{\psi}^i_{\alpha}(x),\;
\Psi^i_{\alpha}(x) \rightarrow \Psi^i_{\alpha}(x) + \psi^i_{\alpha}(x),\,\ldots\,,
\eqno{\rm (B.2)}
$$
where on the left-hand side the functions $A^a_{\mu}$, $\bar{\Psi}^i_{\alpha}$ and $\Psi^i_{\alpha}$ are considered as the classical background fields, and $a^a_{\mu}$, $\bar{\psi}^i_{\alpha}$
and $\psi^i_{\alpha}$ are their quantum fluctuations. Furthermore, substituting (B.2) into (B.1) and retaining only terms which are quadratic in the quantum fields, we have (for simplicity,
the ghost contribution is omitted)
$$
S_{\rm quad}[\,a,\bar{\psi},\psi;A,\bar{\Psi},\Psi] = \!\int\!d^4x\,\Bigl\{\frac{1}{2}\,a^{a{\mu}}(x){\cal D}^{ab}_{\mu\nu}(A,\bar{\Psi},\Psi)a^{b\nu}(x)
$$
$$
+\;a^{a{\mu}}(x)\bar{\cal F}^{\,aj}_{\mu\beta}(A,\bar{\Psi})\psi^j_{\beta}(x)+
\bar{\psi}^i_{\alpha}(x){\cal F}^{\,ib}_{\alpha\nu}(A,\Psi)a^{b\nu}(x)+
\bar{\psi}^i_{\alpha}(x){\cal D}^{\,ij}_{\alpha\beta}(A)\psi^j_{\beta}(x)\Bigr\},
$$
where
\[
\begin{split}
{\cal D}^{ab}_{\mu\nu}(A,\bar{\Psi},\Psi) = g_{\mu\nu}(D_{\lambda}(A)D^{\lambda}(A))^{ab}\,&+ 2\hspace{0.015cm}igF_{\mu\nu}^c(x)(T^c)^{ab}\\
-\,g^{2\,}\bar{\Psi}(x)\bigl(\,g_{\mu\nu}\{t^a\!,t^b\} &+ i\sigma_{\mu\nu}[\,t^a\!,t^b]\bigr)\Psi(x),
\hspace{2cm}
\end{split}
\tag{\rm B.3}
\]
$$
\hspace{0.1cm}
{\cal D}^{\,ij}_{\alpha\beta}(A) = (D_{\mu}(A)D^{\mu}(A))^{ij}\delta_{\alpha\beta} - \frac{1}{2}\,gF^a_{\mu\nu}(x)(t^a)^{ij}(\sigma^{\mu\nu})_{\alpha\beta}
+ m^2\delta^{ij}\delta_{\alpha\beta}
\eqno{\rm (B.4)}
$$
are the kinetic operators of the gauge boson and fermion fluctuations, respectively, and
\[
\begin{split}
&\bar{\cal F}^{\,aj}_{\mu\beta}(A,\bar{\Psi}) = -g\left\{\left[\bar{\Psi}(x)\overleftarrow{D}^{\dagger{\lambda}}(A)\right]\!t^a(2\hspace{0.025cm}ig_{\mu\lambda}\!+\sigma_{\mu\lambda}) +
\left(\bar{\Psi}(x)\,\sigma_{\mu\lambda}t^a\overrightarrow{D}^{\lambda}(A)\right)\right\}^j_{\beta},\\
&{\cal F}^{\,ib}_{\alpha\nu}(A,\Psi) = +\,g\left\{(2\hspace{0.025cm}ig_{\lambda\nu}\!+\sigma_{\lambda\nu})\,t^b\!\left[\overrightarrow{D}^{\lambda}(A)\Psi(x)\right] +
\left(\overleftarrow{D}^{\dagger{\lambda}}(A)\,t^b\sigma_{\lambda\nu}\Psi(x)\right)\right\}^i_{\alpha}
\end{split}
\tag{\rm B.5}
\]
are the mixed contributions to $S_{\rm quad}$. Here, the expressions of the type $\Bigl[\overrightarrow{D}^{\lambda}(A)\Psi(x)\Bigr]$ indicate that the derivative acts only within the square
brackets. By virtue of the fact that we use the second-order formalism for fermions instead of the usual Dirac formalism, the gluon kinetic operator (B.3) contains the term bilinear in
the background fermion fields $\bar{\Psi}^i_{\alpha}$ and $\Psi^i_{\alpha}$. This term creates a new 4-point vertex of interaction when two gauge bosons couple with two fermions \cite{morgan_1995}.\\
\indent
Finally, within the framework of the background gauge fixing technique, the one-loop contribu\-tion to the QCD effective action is
$$
\exp\bigl(i\Gamma^{\rm 1-loop}[\,A,\bar{\Psi},\Psi]\bigr) = \!
\int\!{\cal D}a{\cal D}\psi{\cal D}\bar{\psi}\,\exp\bigl(iS_{\rm quad}[\,a,\bar{\psi},\psi;A,\bar{\Psi},\Psi]\bigr) =
$$
\[
={\rm SDet}^{-1/2}\!
\left(
\begin{array}{cc}
{\cal D}^{ab}_{\mu\nu}(A,\bar{\Psi},\Psi) & \bar{\cal F}^{\,aj}_{\mu\beta}(A,\bar{\Psi}) \\
{\cal F}^{\,ib}_{\alpha\nu}(A,\Psi) & {\cal D}^{\,ij}_{\alpha\beta}(A)
\end{array}
\right),
\]
where the symbol ${\rm SDet}$ denotes a superdeterminant over both gauge bosons and fermions. On the right-hand side we have omitted irrelevant the number factor.

\end{document}